\documentclass[11pt]{article}
\pdfoutput=1
\usepackage{jheppub} 
\usepackage{soul}
\usepackage[normalem]{ulem}

\DeclareFontFamily{U}{wncy}{}
\DeclareFontShape{U}{wncy}{m}{n}{<->wncyr10}{}
\DeclareSymbolFont{mcy}{U}{wncy}{m}{n}
\DeclareMathSymbol{\Sha}{\mathord}{mcy}{"58} 
\def\unit{{\hbox{\kern+.5mm 1\kern-1mm l}}} 

\title{\textbf{$\mathbb{Z}_2$ boundary twist fields and the moduli space of D-branes}}

\abstract{We revisit the boundary conformal field theory of twist fields. Based on the equivalence between twisted bosons on a circle and the orbifold theory at the critical radius, we provide a bosonized representation of boundary twist fields and thus a free field representation of the latter. One advantage of this formulation is that it considerably simplifies the calculation of correlation functions involving twist fields. 
At the same time this also gives access to higher order terms in the operator product expansions of the latter which, in turn, allows to explore the moduli space of marginal deformation of bound states of D-branes. In the process we also generalize some results on correlation functions with excited twist fields.}

\author{Luca Mattiello,}
\author{Ivo Sachs}
\affiliation{Arnold Sommerfeld Center for Theoretical Physics,\\ Ludwig Maximilian University of Munich,\\
Theresienstr. 37, D-80333 M\"unchen, Germany}
\emailAdd{Luca.Mattiello@physik.uni-muenchen.de}
\emailAdd{Ivo.Sachs@physik.uni-muenchen.de}

\begin{document}
\maketitle


\section{Introduction}
Twist fields play an important role in the context of conformal field theory, or more generally, quantum field theory. They are associated to internal symmetries of a QFT; the $\mathbb{Z}_2$ symmetry of the Ising model, for example, is associated to the spin field $\sigma$, the order parameter  field. Similarly, twist fields connect the Ramond and Neveu-Schwarz sectors of a fermion \cite{Ginsparg}. Analogously, there are bosonic twist fields, connecting the Ramond and Neveu-Schwarz sectors for a free boson \cite{Zamolodchikov}.

Twist fields are very important when studying solitons and other non-perturbative effects in string theory that can be described by bound states of D-branes \cite{Polchinski,Polchinski2}, for instance the worldsheet description of black holes \cite{Strominger, Lunin, Giusto} and the reconstruction of the instanton profile in terms of intersecting D-branes \cite{Douglas2, Douglas, Billo}. Generally, the role of twist fields is essential when considering open strings stretched between branes of different dimension, in such a way to have different boundary conditions on the two endpoints; scattering amplitudes contain vertex operators built using twist fields \cite{Hashimoto}.

Another important application of twist fields is in the context of entanglement entropy \cite{Calabrese1,Calabrese}; for example, correlation functions of $\mathbb{Z}_n$ twist fields are connected to the calculation of  the entanglement entropy of an interval.

In this paper we focus on the conformal field theory of $\mathbb{Z}_2$ boundary bosonic twist operators. The essential properties of this theory and the basic correlations functions have been studied extensively, for example in \cite{Frohlich} and references therein. We will review and extend some of these results, including correlation functions involving excited twist fields. Correlators of this kind have been computed in \cite{David:2000yn} and in the context of intersecting D-branes at non-trivial angles \cite{Engberg:1994sd, Anastasopoulos, Pesando}; excited twist fields appear in the vertex operators corresponding to massive stringy excitations. In the case of $\mathbb{Z}_2$ twist, excited twist fields enter in superstring theory, whenever a picture changing of twist vertex operators is needed.

On the other hand, twist fields naturally arise in the context of bulk conformal field theory, since they create the twisted sector of orbifold theories \cite{Dixon}. The connection between orbifolds and other $c=1$ conformal field theories, in particular to the Ashkin-Teller model, has been investigated \cite{Zamolodchikov,Dijkgraaf}. In particular the orbifold theory at $R=\sqrt{2}$ is equivalent to a simple (non-twisted) boson compactified on a circle. This observation will inspire us to define new boundary fields, which we will call \textit{bosonized twist fields}, that have the same local properties as the usual boundary twist fields, but will be argued below to describe an array of Dirichlet sectors, instead of a single one. Since there is a description in which they are mutually local with the bosonic current, this provides a simple derivation of correlation functions of the latter which would be otherwise very complicated. In addition, the  free field representation of our bosonized twist fields gives easy access to their operator product expansion (OPE). This, in turn, leads new insight about the moduli space of bound states of D-branes since we can explore exact marginality of twist field deformations of the boundary conformal field theory. For instance, we find that the modulus corresponding to blowing up a co-dimension $16$ D-brane bound state in bosonic string theory is obstructed for some values of the compactification radius.

The plan of the paper is as follows. In section \ref{sec:free_boson} we review some aspects of the boundary conformal theory of a free boson, and the basic properties of twist fields. In section \ref{sec:bos_twist} we introduce the bosonized twist fields. We explain their relation with the usual twist fields and we argue that they describe an array of Dirichlet sectors. Section \ref{sec:single_insertion} is dedicated to the study of correlation functions on the upper half plane in the presence of two twist fields, using both the operator formalism as well as the properties of twist fields. In section \ref{sec:more_insertions} we consider more twist field insertion, and we present some new explicit results for correlation functions in the presence of four or more twist fields. In section \ref{sec:ordering} we discuss ordering issues when considering the twist fields on the boundary. Section \ref{sec:partition} is devoted to modular invariance of bulk twist field correlation functions and their connection with partition functions on Riemann surfaces. In section \ref{sec:string} we discuss application to string theory, in particular the connection to bound states of intersecting D-branes of different dimensions, which was the original motivation for this work. We also provide the explicit form of some useful correlation functions and we discuss possible marginal deformation of bound states of D-branes in bosonic string theory. In appendix \ref{app:electrostatics} we discuss how correlation functions with twist fields can be derived using the analogy to electrostatics in 2 dimensions. Appendix \ref{app:hyper} contains a discussion about how the calculation of correlation functions is related to the theory of complex functions defined on higher genus Riemann surfaces. In appendix \ref{app:4pt} and \ref{app:jj} we review the derivation of known correlation functions involving four twist fields, and we give the explicit results for other correlators.

\section{Free boson and twist fields}\label{sec:free_boson}
In this section we introduce our normalization conventions for the boundary conformal field theory of a free boson $X$ in one space-time dimension. The bulk action is
\begin{equation}
S[X]=\dfrac{1}{4\pi}\int dz d\bar{z}\; \partial X(z,\bar{z}) \bar{\partial}X(z,\bar{z})\,.
\end{equation}
We use complex coordinates and we split the field in its holomorphic and anti-holomorphic part as $X(z,\bar{z})=X(z)+\bar{X}(\bar{z})$, focusing on the first one. Furthermore, we assume that the domain of $X$ is the upper half plane ($\text{Im}\,z>0$), and the boundary coincides with the real line. Out of the field $X$ one constructs the current
\begin{equation}
j(z)=i\partial X(z)\,,
\end{equation}
which satisfies the OPE
\begin{equation}\label{eq:OPEj3}
j(z)j(w)=\dfrac{1}{(z-w)^2}+\mathrm{reg.}
\end{equation}
The stress-energy tensor is $T(z)=-\frac{1}{2}:\partial X\partial X:(z)=\frac{1}{2}:jj:(z)$, where $:$ $:$ indicates normal ordering. In addition 
\begin{equation}
\widetilde{V}_\alpha(z)=:e^{i\alpha X}:(z)
\end{equation}
are primaries with conformal dimension $h_\alpha=\alpha^2/2$. The OPE among them takes the form
\begin{equation}
\widetilde{V}_\alpha(z)\widetilde{V}_\beta(w)=e^{-\alpha\beta\langle X(z)X(w)\rangle}:e^{i(\alpha+\beta)X}:(w)+\dots=(z-w)^{\alpha\beta}\widetilde{V}_{\alpha+\beta}(w)+\dots
\end{equation}
Finally the OPE with the current $j$ is given by
\begin{equation}\label{eq:VV}
j(z)\widetilde{V}_\alpha(w)=\dfrac{\alpha}{(z-w)}\widetilde{V}_\alpha(w)+\mathrm{reg.}
\end{equation}
Twist fields (also known as  twist operators or boundary changing operators) come into play when dealing with inhomogeneous boundary conditions for the boson $X$, in particular alternating Dirichlet and Neumann boundary conditions. Their operator product expansions with the current $j=i\partial X$ reads (see for example \cite{Zamolodchikov} and \cite{Hashimoto})
\begin{equation}\label{eq:ope_twist}
\begin{split}
&i\partial X(z)\sigma(w)=\dfrac{\sigma'(w)}{(z-w)^{1/2}}+\dots\\
&i\partial X(z)\sigma'(w)=\dfrac{\sigma(w)}{2(z-w)^{3/2}}+\dfrac{2\ \partial\sigma(w)}{(z-w)^{1/2}}+ \dots\\
&\bar{\sigma}(z)\sigma(w)=\dfrac{1}{(z-w)^{1/8}}+\dots \\
\end{split}
\end{equation}
where $\sigma'(w)$ is the excited twist field. In the context of this paper we take this OPE (\ref{eq:ope_twist}) as the defining property of twist fields. Similar relations to the first two hold for the conjugated fields $\bar{\sigma}$ and $\bar{\sigma}'$. $\sigma$ (and its conjugated $\bar{\sigma}$) is a conformal primary of dimension $1/16$, while $\sigma'$ (and $\bar{\sigma}'$) has dimension $9/16$. Notice that the square root branch cut implies that the field $X(z)$ changes sign when the point $z$ is moved around the point where the twist field is inserted. In the following we will always insert twist fields at the boundary of the domain, i.e. on the real line ($z=\bar{z}$). Therefore, the branch cut in the OPE changes the boundary condition from Neumann to Dirichlet (and vice versa). For completeness, we recall that the Neumann intervals on the boundary are characterized by the condition
\begin{equation}
(\partial-\bar{\partial})X(z,\bar{z})\big\vert_{z=\bar{z}}=0\,,
\end{equation}
while on the Dirichlet intervals we have
\begin{equation}
(\partial+\bar{\partial})X(z,\bar{z})\big\vert_{z=\bar{z}}=0\,, \qquad X(z,\bar{z})\big\vert_{z=\bar{z}}=X_0\,.
\end{equation}

\section{Bosonized twist fields}\label{sec:bos_twist}
In general the calculation of correlation functions involving bosonic twist fields is a complicated task (see e.g. \cite{Zamolodchikov,Frohlich}), since these fields are non-local with respect to the boson $X$. For fermions, ghosts and fermionic twist fields (spin fields) it is possible to use a bosonization procedure \cite{Kostelecky} in order to simplify the calculation of correlation functions. In this paper we will apply the same procedure to bosonic twist fields.

\subsection{Orbifold CFTs}
If we consider bulk CFT, one can twist the anti-holomorphic part of the boson as well. Correspondingly we will have another twist field $\sigma(\bar{z})$ satisfying the OPE \eqref{eq:ope_twist} with $\bar{j}=i\bar{\partial}\bar{X}(\bar{z})$. A bulk twist field, twisting the full boson $X(z,\bar{z})$, can be defined by $\sigma(z,\bar{z})=\sigma(z)\sigma(\bar{z})$. In the previous section we have considered $X$ in a non-compact space; if, instead, the boson is compactified on a circle of radius $R$, the insertion of twist fields creates the twisted sector of a symmetric orbifold \cite{Dixon}. Twist fields have the same local properties \eqref{eq:ope_twist} independently of the radius of the orbifold, but correlation functions can be affected by the value of the radius.

To see how a bosonization of twist fields is possible, we recall the classification of conformal field theories at $c=1$ (see, for example, \cite{Ginsparg,Dijkgraaf}).
\begin{figure}[h!]
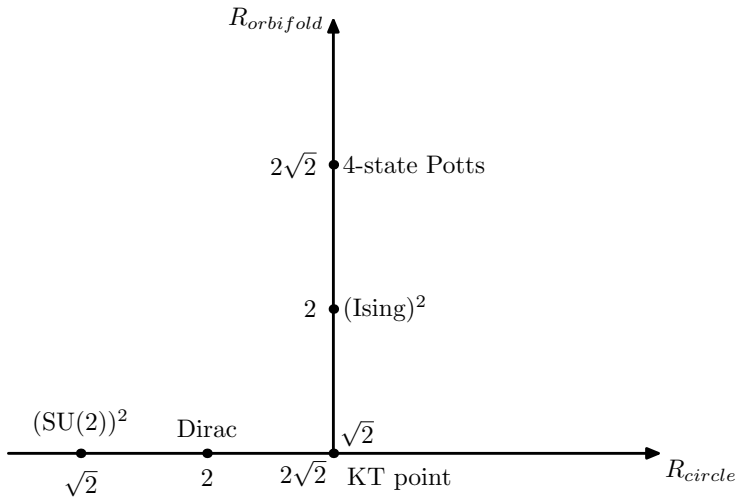

\centering
$\convertMPtoPDF{c1_theories.1}{1.2}{1.2} $
\caption{Classification of conformal field theories at central charge $c=1$.}
\label{fig:c1_theories}
\end{figure}\\
These theories can be divided in two families as in figure \ref{fig:c1_theories}. One describing a boson compactified on a circle $S_1$, and one describing a boson compactified on an orbifold $S_1/\mathbb{Z}_2$. Both of these lines are parametrized by the radius of the circle. Some points on this graph correspond to particular models, for which it is possible to find a description in terms of a boson. It turns out that the two branches in the picture intersect, since the orbifold theory at $R=\sqrt{2}$ is equivalent to the circle theory at $R=2\sqrt{2}$, which corresponds to the continuum limit of the XY-model at the Kosterlitz-Thouless point \cite{Kadanoff}. This duality is valid also at the level of boundary CFT; it has been shown that the two bulk CFT admit the same boundary conditions and boundary operators or, in string theory language, the same set of D-branes \cite{Gaberdiel:2008zb}. This leads us to the idea of bosonizing boundary changing  operators (the bosonized twist fields) in terms of another boson compactified on $S^1$. 

\subsection{\texorpdfstring{$\mathfrak{su}(2)$  Ka\v{c}-Moody algebra}{su(2) Kac-Moody algebra}}
Let us consider first the holomorphic part of the free boson $X(z)$; let us consider the primary operators $\widetilde{V}_\alpha(z)$, in particular the ones with $\alpha=\pm\sqrt{2}$. These are allowed operators when the boson is compactified on a circle at the self-dual radius. Out of these two operators we can construct two other currents, namely:
\begin{equation}
j^1(z)=\dfrac{1}{\sqrt{2}}\left(\widetilde{V}_{\sqrt{2}}(z)+\widetilde{V}_{-\sqrt{2}}(z)\right)\,,\qquad\quad j^2(z)=\dfrac{i}{\sqrt{2}}\left(\widetilde{V}_{\sqrt{2}}(z)-\widetilde{V}_{-\sqrt{2}}(z)\right)\,.
\end{equation}
The three currents $j^1$, $j^2$ and $j^3=j=i\partial X$ constitute a $\mathfrak{su}(2)$ Ka\v{c}-Moody algebra.

To continue we perform a change of basis by introducing a new free chiral boson $\Omega(z)$, satisfying the same OPE
\begin{equation}
\Omega(z)\Omega(w)\sim -\log(z-w)\,,
\end{equation}
as the chiral field $X(z)$. Out of the boson $\Omega$ we can construct three currents $J^i(z)$ ($i=1,2,3$), analogously to the currents $j^i$ constructed out of $X$. We then express $\partial X$ in terms of $\Omega$, identifying
\begin{equation}\label{eq:current}
i\partial X(z)=j^3(z)\equiv J^2(z)=\dfrac{i}{\sqrt{2}}\left(:e^{i\sqrt{2}\Omega}:(z)-:e^{-i\sqrt{2}\Omega}:(z)\right)\,.
\end{equation}
This change of basis is equivalent to a rotation in the three-dimensional space generated by the three currents of the $\mathfrak{su}(2)$  Ka\v{c}-Moody algebra (cfr. \cite{Ginsparg}). For consistency we also impose the identifications
\begin{equation}\label{eq:dO}
\begin{split}
i\partial \Omega(z)=J^3(z)\equiv &j^1(z)=\dfrac{1}{\sqrt{2}}\left(:e^{i\sqrt{2}X}:(z)+:e^{-i\sqrt{2}X}:(z)\right)\,,\\
\dfrac{1}{\sqrt{2}}\left(:e^{i\sqrt{2}\Omega}:(z)+:e^{-i\sqrt{2}\Omega}:(z)\right)&=J^1(z)\equiv j^2(z)=\dfrac{i}{\sqrt{2}}\left(:e^{i\sqrt{2}X}:(z)-:e^{-i\sqrt{2}X}:(z)\right)\,.
\end{split}
\end{equation}
Once we have done this rotation, we have a description of the CFT of a free boson in another basis. This may not seem convenient, since the conformal primaries $:e^{i\alpha X}:$ do not have a local description in terms of $\Omega$ in this new picture. However, this rotation allows us to identify new primaries which do not have a local description in terms of $X$. Namely they are the primaries $V_\alpha(z)=:e^{i\alpha\Omega}:(z)$, and the bosonized twist fields will be among them.

The same procedure can be done for the anti-holomorphic part of the boson, defining an anti-chiral field $\bar{\Omega}$. Notice that $\Omega(z,\bar{z})=\Omega(z)+\bar{\Omega}(\bar{z})$, as a functional of $X(z,\bar{z})=X(z)+\bar{X}(\bar{z})$, will be periodic under a shift of $2\pi\sqrt{2} n$ ($n\in \mathbb{Z}$), which means that also $\Omega$ is compactified on a circle at the self-dual radius. Let us now consider the boson $X(z,\bar{z})$ compactified on an orbifold at the self-dual radius, with the $\mathbb{Z}_2$ transformation defined by $X\rightarrow -X$. The boson $\Omega(z,\bar{z})$ should be unaffected by this transformation; this can be achieved if $\Omega(z,\bar{z})$ is compactified on a circle with half the radius, namely $R=1/\sqrt{2}$. Notice that this is consistent with the identifications \eqref{eq:current} and \eqref{eq:dO}. Furthermore, T-duality implies that the circle theory at $R=1/\sqrt{2}$ is in turn equivalent to a circle theory at radius $R'=2/R=2\sqrt{2}$, in accordance with what depicted in figure \ref{fig:c1_theories}.  

\subsection{Boundary conditions and bosonized twist fields}
Let us now consider a boundary CFT, with the chiral and anti-chiral part of the boson $X$ related by boundary conditions. If we define the chiral and anti-chiral part of a boson $\Omega$ as in the previous subsection, we can deduce the boundary conditions in terms of $\Omega$. For this we note that a change of sign in $\partial X(z)$, which one needs in order to interchange Dirichlet and Neumann boundary conditions, can be achieved by shifting the chiral field $\Omega(z)$ by $\pi/\sqrt{2}$; therefore Neumann and Dirichlet boundary conditions for $X$ (on the real line $z=\bar{z}$) correspond to 
\begin{equation}
\begin{split}
&\text{Neumann:} \qquad \Omega(z)=\bar{\Omega}(\bar{z})\,,\\
&\text{Dirichlet:  } \qquad \Omega(z)=\bar{\Omega}(\bar{z})+\dfrac{\pi}{\sqrt{2}}\,.
\end{split}
\end{equation}
These boundary conditions may look unfamiliar from the point of view of the $\Omega$ boundary conformal field theory. However, it is not hard to see that they are conformal as it must be since they correspond to the usual Neumann and Dirichlet boundary conditions for the boson $X$.

Among the primaries $V_\alpha$, defined in terms of $\Omega$, there are some that have the same local properties \eqref{eq:ope_twist} as  twist fields. Indeed, consider the two primaries $\sigma_B=V_{\sqrt{2}/4}$ and $\bar{\sigma}_B=V_{-\sqrt{2}/4}$, both with conformal dimension $1/16$. $\sigma_B$ will be called bosonized twist field, and $\bar{\sigma}_B$ is its conjugated field. Moreover, we  identify the excited  bosonized twist field $\sigma'_B=-\frac{i}{\sqrt{2}}V_{-3\sqrt{2}/4}$ and its conjugated $\bar{\sigma}'_B=\frac{i}{\sqrt{2}}V_{3\sqrt{2}/4}$. 
Given these definitions, the bosonized twist fields satisfy the following OPE's with the current $j=i\partial X$:
\begin{equation}
\begin{split}
&i\partial X(z)\sigma_B(w)=\dfrac{\sigma'_B(w)}{(z-w)^{1/2}}+\dots\\
&i\partial X(z)\sigma'_B(w)=\dfrac{\sigma_B(w)}{2(z-w)^{3/2}}+\dfrac{2\ \partial\sigma_B(w)}{(z-w)^{1/2}}+ \dots\,,
\end{split}
\end{equation}
which are identical to \eqref{eq:ope_twist}, at least for the most divergent terms. The description in terms of $\Omega$ allows to treat the bosonized twist fields and the current $\partial X$ in the same way, having a free field representation for all of them at the same time.

The way we interpret these (boundary) twist fields is the following: usually twist fields relate the Neumann sector ($(\partial-\bar{\partial}) X=0$) to the Dirichlet sector $X(z,\bar{z})=X_0$, where $X_0$ is some value for the boundary condition, and this information is encoded in the twist fields. The bosonized version $\sigma_B=V_{\sqrt{2}/4}$, however, cannot provide the information about $X_0$; furthermore, the periodicity property $X(z,\bar{z})\sim X(z,\bar{z})+2\pi\sqrt{2} n$, which is necessary for the definition of the boson $\Omega$, suggests that these twist fields describe the superposition of different Dirichlet sectors with boundary conditions $X_0^n=2\pi\sqrt{2} n$ ($n\in \mathbb{Z}$). We will give more evidence in favour of this interpretation in the following sections.\footnote{We would like to thank Carlo Maccaferri for sharing his insight with us on this point.}

\section{Single twist field insertion}\label{sec:single_insertion}
Let us begin by reviewing some basic facts about the  boundary CFT of a free boson $X$ in the presence of a single Dirichlet sector, i.e. in the presence of a $\bar{\sigma}$-$\sigma$ pair, and compare it with the bosonized version $\bar{\sigma}_B$-$\sigma_B$. The presence of two twist fields is needed, because one of them, say $\sigma$ changes the boundary condition from Dirichlet to Neumann, while $\bar{\sigma}$ changes from Neumann to Dirichlet. In the bosonized language, the presence of both a twist field and its conjugated is necessary, since the integration over the zero mode of $\Omega$ implies that the sum of all the exponents $\alpha_i$ in $\langle \prod_i V_{\alpha_i}\rangle$ has to be zero. Hence correlation functions with just one $\sigma$ (or just one $\sigma_B$) would vanish.

\subsection{Free boson with anti-periodic boundary condition}
We start with the operator formalism for Dirichlet and Neumann boundary conditions on the positive and negative real line respectively. In this sector the mode expansion for the two currents $j$ and $\bar j$ are 
\begin{equation}\label{eq:exp_R}
\begin{split}
&j(z)=i\partial X(z)=\sum_{r\in\mathbb{Z}+\frac{1}{2}} j_r z^{-r-1}\,,\\
&\bar{j}(\bar{z})=i\bar{\partial} \bar{X}(\bar{z})=-\sum_{r\in\mathbb{Z}+\frac{1}{2}} j_r \bar{z}^{-r-1}\,.
\end{split}
\end{equation}
where the modes satisfy the commutation relation $[j_r,j_s]=r\delta_{r+s}$. A branch cut is present in the complex plane, extending from $0$ to $-\infty$. Equation \eqref{eq:exp_R} defines a boson in the Ramond sector, instead of the usual integer mode expansion, which corresponds to the Neveu-Schwarz sector. The twist vacuum (and its dual) are related to the NS vacuum through
\begin{equation}\label{eq:vacuum}
\begin{split}
&\vert\sigma\rangle=\lim_{z\rightarrow 0}\sigma(z)\vert 0\rangle\,,\\
&\langle\sigma\vert=\lim_{z\rightarrow \infty}z^{1/8}\langle 0\vert\bar{\sigma}(z)\,.
\end{split}
\end{equation}
These states are normalized,  $\langle\sigma\vert\sigma\rangle=1$ and the modes $j_r$ are creation and annihilation operators for $\vert\sigma\rangle$:
\begin{equation}\label{eq:cre_ann}
\begin{split}
&j_r\vert\sigma\rangle=0\,,\qquad r\geq 1/2\,,\\
&\langle\sigma\vert j_r=0\,,\qquad r\leq -1/2\,.
\end{split}
\end{equation}
Other states in the Ramond sector can be obtained applying creation operators to the vacuum. For example we can define the state $\vert\sigma'\rangle$ and its conjugated as
\begin{equation}
\vert\sigma'\rangle=j_{-1/2}\vert\sigma\rangle\,,\qquad \langle\sigma'\vert=\langle\sigma\vert j_{1/2}\,.
\end{equation}
These states are associated (through the state-operator correspondence) to the excited twist fields.

The expansion \eqref{eq:exp_R} is not defined on the negative real axis; we can formally solve the problem introducing a new current $\mathsf{j}(z)$, defined on the whole complex plane as
\begin{equation}
\mathsf{j}(z)=\sum_{r\in\mathbb{Z}+\frac{1}{2}} j_r z^{-r-1}\,.
\end{equation}
This means that we are identifying $\mathsf{j}(z)=j(z)$ on the upper half plane, and $\mathsf{j}(z)=-\bar{j}(z)$ on the lower half plane. This new current is naturally defined on the two-fold branched cover of the complex plane. Therefore, the problem of computing correlation functions of $j(z)$ on the upper half plane with anti-periodic boundary conditions is equivalent to that  of correlation functions of $\mathsf{j}(z)$ on the two-fold cover of the complex plane, restricting the result to $\text{Im}\,z>0$ on the first sheet. 

\subsection{Normal ordering}\label{subsec:normal_ordering}
In the case of a boson in the Ramond sector we have two useful definitions of normal ordering which do not coincide (see e.g. \cite{DiFrancesco}). The first one (indicated with $N(\;)$) arises from the operator product expansion, the second one (indicated with $:\;:$) is a prescription on the order of annihilation and creation modes. Let us consider the OPE of two currents $j$; the normal ordered product $N(jj)$ is the finite term of the expansion, i.e.
\begin{equation}
j(z)j(w)=\dfrac{1}{(z-w)^2}+N(jj)(w)+\dots
\end{equation}
which gives the familiar Sugawara construction of the energy momentum tensor, $T(w)=\frac{1}{2}N(jj)(w)$. It is related to the creation-annihilation normal ordering through
\begin{equation}
T(w)=\dfrac{1}{2}N(jj)(w)=\dfrac{1}{2}:jj:(w)+\dfrac{1}{16w^2}\,,
\end{equation}
so that
\begin{equation}
\langle\sigma\vert T(w)\vert\sigma\rangle=\dfrac{1}{16w^2}\,.
\end{equation}
Consequently, the Laurent modes $L_m$ of $T$ are
\begin{equation}
\begin{split}
&L_m=\dfrac{1}{2}\sum_{r\in\mathbb{Z}+1/2}j_r\,\,j_{m-r}\,,\qquad (m\neq 0)\\
&L_0=\dfrac{1}{16}+\sum_{r=1/2}^\infty j_{-r}\,\,j_r\,.
\end{split}
\end{equation}
where the shift by $1/16$ is just the conformal weight of the field $\sigma$. 

\subsection{Correlation functions with two twist fields}\label{sec:corr_2_twsit}
The CFT of a boson with anti-periodic boundary conditions has been studied intensively (see e.g. \cite{Corrigan:1975sn, Ginsparg}). Some correlation functions can be derived simply by means of the mode expansion, and using the property \eqref{eq:cre_ann}. For example one obtains:
\begin{equation}\label{eq:OPE_R}
\begin{split}
\langle\sigma\vert\mathsf{j}(&z)\mathsf{j}(w)\vert\sigma\rangle=\dfrac{1}{2}\dfrac{\left(\sqrt{\frac{z}{w}}+\sqrt{\frac{w}{z}}\right)}{(z-w)^2}\,,\\
&\langle\sigma\vert\mathsf{j}(z)\vert\sigma'\rangle=\dfrac{1}{2z^{3/2}}\,,\\
&\langle\sigma'\vert\mathsf{j}(z)\vert\sigma\rangle=\dfrac{1}{2z^{1/2}}\,.
\end{split}
\end{equation}
Via the operator state correspondence, \eqref{eq:OPE_R} should be interpreted in terms of correlation functions in the presence of two twist fields (or excited twist fields) at the ramification points $0$ and $\infty$. More generally,
\begin{equation}\label{eq:corr_2twist}
\begin{split}
\langle \bar{\sigma}(z_1) j(z_2) j(z_3)&\sigma(z_4)\rangle=\dfrac{1}{2}\dfrac{1}{(z_{41})^{1/8}(z_{32})^2}\left(\sqrt{\dfrac{z_{31}z_{42}}{z_{21}z_{43}}}+\sqrt{\dfrac{z_{21}z_{43}}{z_{31}z_{42}}}\right)\,,\\
&\langle \bar{\sigma}(z_1) j(z_2) \sigma'(z_3)\rangle=\dfrac{z_{31}^{3/8}}{2z_{21}^{1/2}z_{32}^{3/2}}\,,\\
&\langle \bar{\sigma}'(z_1) j(z_2) \sigma(z_3)\rangle=\dfrac{z_{31}^{3/8}}{2z_{21}^{3/2}z_{32}^{1/2}}\,.
\end{split}
\end{equation}
where $z_{ij}=z_i-z_j$. The last two correlation functions can also be derived taking appropriate limits of the first one, and using the OPE \eqref{eq:ope_twist} defining excited twist fields. For example
\begin{equation}
\langle \bar{\sigma}(z_1) j(z_2) \sigma'(z_4)\rangle=\lim_{z_3\rightarrow z_4}\sqrt{z_3-z_4}\langle \bar{\sigma}(z_1) j(z_2) j(z_3)\sigma(z_4)\rangle\,.
\end{equation} 
Other correlation functions are well known, including the primaries 
\begin{equation}
\psi_\alpha(z)\propto N(e^{i\alpha X})(z)\propto \dfrac{:e^{i\alpha X}(z):}{z^{\alpha^2/2}}\,,
\end{equation}
with conformal weight $h_\alpha=\alpha^2/2$, and OPE
\begin{equation}
\begin{split}
&T(z)\psi_\alpha(w)=\dfrac{h_\alpha}{(z-w)^2}\psi_\alpha(w)+\dfrac{1}{z-w}\partial_w\psi_\alpha(w)+\dots\\
&j(z)\psi_\alpha(w)=\dfrac{\alpha}{z-w}\sqrt{\dfrac{w}{z}}\psi_\alpha(w)+\dots
\end{split}
\end{equation}
The expansion over half-integer modes is a powerful tool that allows computing correlators on the Ramond vacuum, i.e. of the form $\langle\sigma\vert\phi_1(z_1)\phi_2(z_2)\dots\vert\sigma\rangle$, even if some of the fields are of the form $\psi_\alpha$. From these results one can derive the corresponding correlation function with two twist fields on the vacuum $\vert 0\rangle$, using the definitions \eqref{eq:cre_ann}. For example we can compute (see \cite{Frohlich} and \cite{Mukhopadhyay}):
\begin{equation}\label{eq:ssexp}
\langle\sigma\vert\psi_\alpha(z)\vert\sigma\rangle=\dfrac{e^{i\alpha x_0}}{z^{\alpha^2/2}}\,,
\end{equation}
where a particular normalization for $\psi_\alpha$ has been chosen. Here  $x_0$ is the zero mode of the chiral boson $X(z)$; we denote the zero mode of $X(z,\bar{z})$ by $X_0=x_0+\bar{x}_0$, where $\bar{x}_0$ is the zero mode of the anti-chiral part. Analogously:
\begin{equation}\label{eq:psipsi2twists}
\langle\sigma\vert\psi_\alpha(z)\psi_\beta(w)\vert\sigma\rangle= \dfrac{e^{i(\alpha+\beta)x_0}}{z^{\alpha^2/2}w^{\beta^2/2}}\left(\dfrac{1-\sqrt{w/z}}{1+\sqrt{w/z}}\right)^{\alpha\beta}\,.
\end{equation}
Using conformal symmetry we can derive the correlators involving two twist fields:
\begin{equation}
\begin{split}
&\langle\bar{\sigma}(z_1)\psi_\alpha(z_2)\sigma(z_3)\rangle=\dfrac{e^{i\alpha x_0}}{z_{21}^{\alpha^2/2}z_{32}^{\alpha^2/2}z_{31}^{1/8-\alpha^2/2}}\,,\\
\langle\bar{\sigma}(z_1)\psi_\alpha(z_2)\psi_\beta&(z_3)\sigma(z_4)\rangle= \dfrac{e^{i(\alpha+\beta)x_0}}{(z_{21}z_{42})^{\alpha^2/2}(z_{31}z_{43})^{\beta^2/2}z_{41}^{1/8-\alpha^2/2-\beta^2/2}}\left(\dfrac{1-\sqrt{\eta}}{1+\sqrt{\eta}}\right)^{\alpha\beta}\,,
\end{split}
\end{equation}
where $\eta$ is the conformal ratio $\eta=\frac{z_{21}z_{43}}{z_{31}z_{42}}$. We note, in passing, that all these results can be derived in another way, using the analogy to electrostatics in two dimensions (see appendix \ref{app:electrostatics}). Excited twist fields can be also considered; for example, one can derive the following correlation functions (for insertions of the twist fields at generic positions):
\begin{equation}
\begin{split}
\langle\bar{\sigma}(z_1)\psi_\alpha(z_2)\sigma'(z_3)\rangle&=\dfrac{-\alpha e^{i\alpha x_0}}{z_{21}^{\alpha^2/2-1/2}z_{32}^{\alpha^2/2+1/2}z_{31}^{5/8-\alpha^2/2}}\,,\\
\langle\bar{\sigma}'(z_1)\psi_\alpha(z_2)\sigma(z_3)\rangle&=\dfrac{\alpha e^{i\alpha x_0}}{z_{21}^{\alpha^2/2+1/2}z_{32}^{\alpha^2/2-1/2}z_{31}^{5/8-\alpha^2/2}}\,,\\
\langle\bar{\sigma}'(z_1)\psi_\alpha(z_2)\sigma'(z_3)\rangle&=\dfrac{-\alpha^2 e^{i\alpha x_0}}{z_{21}^{\alpha^2/2}z_{32}^{\alpha^2/2}z_{31}^{9/8-\alpha^2/2}}\,.
\end{split}
\end{equation}
From the correlation functions of twist fields with operators $\psi_\alpha$ we can see that the operator product expansion of two twist fields must contain all these primaries. Therefore we can guess that
\begin{equation}\label{dope}
\bar{\sigma}(z)\sigma(w)=\int d\alpha\, \dfrac{e^{-i\alpha x_0}}{(z-w)^{1/8-\alpha^2/2}}\,\psi_\alpha(w)+\dots\,,
\end{equation}
where $x_0$ is the Dirichlet boundary condition of the interval between the insertion of the two twist fields. The rest of the OPE contains descendants of $\psi_\alpha$, but can in principle contain also other primaries. If the boson is compactified, the values of $\alpha$ are restricted, and the integral becomes a sum. For example, in the case of bosonized twist fields, $\alpha$ must be a multiple of $\sqrt{2}$. This can be derived also in the $\Omega$ picture, since
\begin{equation}
\bar{\sigma}_B(z)\sigma_B(w)=\dfrac{1}{(z-w)^{1/8}}\left(1-\dfrac{\sqrt{2}}{4}(z-w)i\partial\Omega(w)+\dots\right)\,,
\end{equation}
and $\Omega$ is expressed in terms of exponential operators through \eqref{eq:dO}.

\subsection{Correlation functions with two bosonized twist fields}
The same correlation functions considered above can be computed for the bosonized twist fields since the current $j=i\partial X$ and the bosonized twist fields have both a local description in terms of the boson $\Omega$. Correlation functions involving only $j$ and bosonized twist fields are straightforward; for the explicit calculation we use 
\begin{equation}\label{eq:corr_sphere}
\langle V_{\alpha_1}(z_1)\dots V_{\alpha_n}(z_n)\rangle=\prod_{i<j}(z_{ij})^{\alpha_i \alpha_j}\,\,\delta\left(\sum_{i=1}^n \alpha_i\right)\,.
\end{equation}
From this one obtains
\begin{equation}\label{eq:corr_2twist_boson}
\begin{split}
\langle \bar{\sigma}_B(z_1) j(z_2) j(z_3)&\sigma_B	(z_4)\rangle=\dfrac{1}{2}\dfrac{1}{(z_{41})^{1/8}(z_{32})^2}\left(\sqrt{\dfrac{z_{31}z_{42}}{z_{21}z_{43}}}+\sqrt{\dfrac{z_{21}z_{43}}{z_{31}z_{42}}}\right)\,,\\
&\langle \bar{\sigma}_B(z_1) j(z_2) \sigma'_B(z_3)\rangle=\dfrac{z_{31}^{3/8}}{2z_{21}^{1/2}z_{32}^{3/2}}\,,\\
&\langle \bar{\sigma}'_B(z_1) j(z_2) \sigma_B(z_3)\rangle=\dfrac{z_{31}^{3/8}}{2z_{21}^{3/2}z_{32}^{1/2}}\,,
\end{split}
\end{equation}
in agreement with \eqref{eq:corr_2twist}. This is not surprising since the zero mode of $X$ does not appear in the current $j=i\partial X$.

When fields like $\psi_\alpha$ are present, correlation functions depend explicitly on the particular boundary condition $x_0$. Therefore, we do expect them to be different when the bosonized twist fields $\sigma_B$ and $\bar{\sigma}_B$ are used instead of $\sigma$ and $\bar{\sigma}$. As explained before, however, we want to interpret the result in terms of a superposition of different sectors with boundary conditions $x_0=\sqrt{2}\pi n$ (considering only the holomorphic part of $X$).  We thus assume that
\begin{equation}\label{eq:Ansatz}
\langle\bar{\sigma}_B(z_1)\psi_\alpha(z_2)\sigma_B(z_3)\rangle=\sum_{n\in\mathbb{Z}}\langle\bar{\sigma}(z_1)\psi_\alpha(z_2)\sigma(z_3)\rangle_{x_0=\sqrt{2}\pi n}\,,
\end{equation}
and check this hypothesis. Since the sum is over an infinite number of Dirichlet sectors, the result must be normalized, in order to match with the normalization of two-point function $\langle\bar{\sigma}(z_1)\sigma(z_3)\rangle=z_{13}^{-1/8}$. The right hand side  of \eqref{eq:Ansatz} is straightforward to compute, and involves the Dirac comb
\begin{equation}
\Sha_T(t)=\dfrac{1}{T}\sum_{n\in\mathbb{Z}}e^{2\pi i n \frac{t}{T}}=\sum_{k\in\mathbb{Z}}\delta(t-kT)\,.
\end{equation}
The final result for the right hand side of \eqref{eq:Ansatz} is
\begin{equation}\label{eq:Dirac_comb}
\dfrac{\sqrt{2}}{z_{13}^{1/8}}\left(\dfrac{z_{13}}{z_{12}z_{23}}\right)^{\alpha^2/2}\Sha_{\sqrt{2}}(\alpha)=\dfrac{\sqrt{2}}{z_{13}^{1/8}}\left(\dfrac{z_{13}}{z_{12}z_{23}}\right)^{\alpha^2/2}\left(\delta(\alpha)+\delta(\alpha+\sqrt{2})+\delta(\alpha-\sqrt{2})+\dots\right)\,.
\end{equation}
The calculation of the left hand side  is more involved, since $\sigma_B$ and $\bar{\sigma}_B$ are naturally written in terms of $\Omega$, while $\psi_\alpha$ is not local with respect to it. However, we can rewrite the combination $\bar{\sigma}_B(z_1)\sigma_B(z_3)$ as
\begin{equation}
\bar{\sigma}_B(z_1)\sigma_B(z_3)=V_{-\sqrt{2}/4}(z_1)V_{\sqrt{2}/4}(z_3)=\dfrac{1}{z_{31}^{1/8}}\exp\left(\dfrac{\sqrt{2}}{4}\int_{z_3}^{z_1}i\partial\Omega(z)dz\right)\,
\end{equation}
and express $\partial\Omega$ in terms of $X$ using \eqref{eq:dO}. This gives a path integral over $X$,
\begin{equation}
\langle\bar{\sigma}_B(z_1)\psi_\alpha(z_2)\sigma_B(z_3)\rangle=\dfrac{1}{Z}\int[dX] \left(\bar{\sigma}_B(z_1)\psi_\alpha(z_2)\sigma_B(z_3)\right) e^{-S[X]}\,.
\end{equation}
We can split the  integral as $\int[dX]=\int dx_0\int [dX_\perp]$, where $x_0$ represents the zero mode of $X$. The periodicity properties of $\partial\Omega$ imply that the $\int dx_0$ integral is of the form
%
\begin{equation}
\int dx_0\, e^{i\alpha x_0}f(x_0)\,,
\end{equation}
where $e^{i\alpha x_0}$ accounts for the zero mode in $\psi_\alpha$ and $f$ is a periodic function $f(x)=f(x+\sqrt{2}\pi)$. This integral can be rewritten as
\begin{equation}
 \sum_{n\in\mathbb{Z}} e^{i\alpha n\sqrt{2}\pi}\int_0^{\sqrt{2}\pi}dx_0\, f(x_0) e^{i\alpha x_0}\,.
\end{equation}
Using the definition of Dirac comb, we then notice that
\begin{equation}
\langle\bar{\sigma}_B(z_1)\psi_\alpha(z_2)\sigma_B(z_3)\rangle\propto \Sha_{\sqrt{2}}(\alpha)=\left(\delta(\alpha)+\delta(\alpha+\sqrt{2})+\delta(\alpha-\sqrt{2})+\dots\right)\,,
\end{equation}
in agreement with \eqref{eq:Dirac_comb}. Furthermore, the dependence on the positions $z_1$, $z_2$ and $z_3$ is fixed by conformal invariance, hence it must coincide with the one in \eqref{eq:Dirac_comb} up to normalisation.

\section{More twist field insertions}\label{sec:more_insertions}
With four or more twist fields the situation is more complicated, for three reasons. First, the operator formalism reviewed in section \ref{eq:exp_R} is not applicable. The second reason is that we have two or more cuts on the complex plane where the fields are defined; this means that the worldsheet is now effectively a hyperelliptic surface with genus $g>0$ \cite{Hamidi:1986vh, Frohlich}. Finally, using the electrostatic analogy for finding correlation functions on the upper half plane is still possible, but an explicit expression for the Green's function with appropriate boundary conditions is known only in integral form. In appendix \ref{app:hyper} we review the connection between twist fields insertions and hyperelliptic surfaces, in particular in the case of four twist fields.

\subsection{Correlation functions with four twist fields}
Let us now consider correlation functions involving four twist fields; many of them are already known, and have been derived solving systems of differential equations, similar to the Knizhnik-Zamolodchikov equations (see \cite{Zamolodchikov,Frohlich,Gava:1997jt} and also \cite{Cvetic:2003ch,Abel:2003vv} for parallel results in the context of D-branes at angles). We review here some of these results, and we extend them to the corresponding correlation functions involving the bosonized version of the twist fields. Let us consider first of all the correlation function of four twist fields, namely
\begin{equation}
\langle\bar{\sigma}(z_1)\sigma(z_2)\bar{\sigma}(z_3)\sigma(z_4)\rangle\,.
\end{equation}
This correlation function is well known in the literature, and was computed for example in \cite{Zamolodchikov}; the detailed derivation can be found in appendix \ref{app:4pt}. It is important to notice that in the presence of four twist fields there are two Dirichlet intervals on the boundary. The boson $X(z,\bar{z})$ can in principle have different boundary conditions $X=X_0^i$ ($i=1,2$) on the two intervals. Adapting the result of \cite{Zamolodchikov} to our notations, the correlation function is
\begin{equation}
\langle\bar{\sigma}(z_1)\sigma(z_2)\bar{\sigma}(z_3)\sigma(z_4)\rangle=\left(\dfrac{z_{31}z_{42}}{z_{21}z_{41}z_{32}z_{43}}\right)^{1/8}\sqrt{\dfrac{\pi}{2K(\eta)}}\exp\left(\frac{i}{8\pi}(X_0^1-X_0^2)^2\tau\right)\,,
\end{equation}
where the conformal ratio $\eta$ is given by $\eta=z_{43}z_{21}/(z_{42}z_{31})$, and $K(\eta)$ is the complete elliptic integral of the first kind. In order to derive this four-point function one encounter other correlation functions, namely $\langle j(w)\bar{\sigma}(z_1)\sigma(z_2)\bar{\sigma}(z_3)\sigma(z_4)\rangle$, $\langle\bar{\sigma}(z_1)\sigma'(z_2)\bar{\sigma}(z_3)\sigma(z_4)\rangle$ and $\langle j(w)\bar{\sigma}(z_1)\sigma'(z_2)\bar{\sigma}(z_3)\sigma(z_4)\rangle$; explicit expressions are given in appendix \ref{app:4pt}.

Another important correlation function that can be computed is the one involving two currents $j$ and four twist fields. The result is known (see \cite{Dixon,Gava:1997jt}) when the difference of the two boundary conditions $\delta=X_0^1-X_0^2$ is zero. In appendix \ref{app:jj} we generalize to the case $\delta\neq 0$, the result being
\begin{equation}\label{eq:jjssss}
\begin{split}
\langle j(z)j(w)&\bar{\sigma}(z_1)\sigma(z_2)\bar{\sigma}(z_3)\sigma(z_4)\rangle=\\
=&\dfrac{G(z_i)}{2(z-w)^2}\left[\sqrt{\dfrac{(z-z_1)(w-z_2)(z-z_3)(w-z_4)}{(w-z_1)(z-z_2)(w-z_3)(z-z_4)}}+(z \leftrightarrow w)\right]+\\
&+\dfrac{\sqrt{2\pi}}{\sqrt{P(z)P(w)}}\left(\dfrac{z_{31}z_{42}}{z_{21}z_{41}z_{32}z_{43}}\right)^{-7/8}\partial_\eta\left[\dfrac{1}{\sqrt{K(\eta)}}\exp\left(\frac{i\delta^2}{8\pi}\tau(\eta)\right)\right]\,,
\end{split}
\end{equation}
where $G(z_i)$ is the four-point function $G(z_i)=\langle\bar{\sigma}(z_1)\sigma(z_2)\bar{\sigma}(z_3)\sigma(z_4)\rangle$.

Starting with the correlation function with two currents one can easily obtain other correlation functions involving excited twist fields. It is sufficient to consider the limit when one of the currents approaches a twist fields, and use the corresponding OPE, as done in section \ref{sec:single_insertion}. In particular, we derive explicit results for $\langle\bar{\sigma}'(z_1)\sigma'(z_2)\bar{\sigma}(z_3)\sigma(z_4)\rangle$ and $\langle\bar{\sigma}'(z_1)\sigma(z_2)\bar{\sigma}'(z_3)\sigma(z_4)\rangle$ in appendix \ref{app:jj}. Similar and other correlation functions involving excited twist fields can be found in \cite{David:2000yn}, and in \cite{Anastasopoulos2} for the case of twist fields connecting D-branes at different angles.

\subsection{Correlation functions with four bosonized twist fields}
We now compare the above results to the case of four bosonized twist fields. The calculation of the four twist correlator  is straightforward using \eqref{eq:corr_sphere}:
\begin{equation}\label{eq:4ptB}
\langle\bar{\sigma}_B(z_1)\sigma_B(z_2)\bar{\sigma}_B(z_3)\sigma_B(z_4)\rangle=\left(\dfrac{z_{31}z_{42}}{z_{21}z_{41}z_{32}z_{43}}\right)^{1/8}\,.
\end{equation}
This correlation function should represent a double array of Dirichlet sectors, whose boundary conditions are separated by $2\sqrt{2}\pi$. This is because each pair $\bar{\sigma}_B$-$\sigma_B$ connects the Neumann sector to the array; therefore a sum over the array has to be performed for both pairs. Thus we should compare \eqref{eq:4ptB} with the quantity
\begin{equation}\label{eq:4pt_sum}
\sum_{a,b\in\mathbb{Z}}\left(\dfrac{z_{31}z_{42}}{z_{21}z_{41}z_{32}z_{43}}\right)^{1/8}\sqrt{\dfrac{\pi}{2K(\eta)}}\exp\left(\frac{i}{8\pi}(2\sqrt{2}\pi(a-b))^2\tau\right)\,.
\end{equation}
The sum is infinite but will give a finite result after dividing by the two-point function\\$\sum_{n\in\mathbb{Z}} \langle\bar{\sigma}(z_1)\sigma(z_2)\rangle_{X_0=2\sqrt{2}\pi n}$, which is the correct normalization of correlation functions. Using the Jacobi theta function $\vartheta_3$, which satisfies
\begin{equation}
\vartheta_3(0;\tau)=\sum_{n\in\mathbb{Z}}e^{i\pi\tau n^2}=\sqrt{\dfrac{2K(\eta)}{\pi}}\,,
\end{equation}
we notice that \eqref{eq:4pt_sum} is equal to \eqref{eq:4ptB}.

We can proceed in an analogous way for the correlation function with two currents and four twist fields. Using the bosonized expression of $j$ one easily derives
\begin{equation}\label{eq:jjssssB}
\begin{split}
&\langle j(z)j(w)\bar{\sigma}_B(z_1)\sigma(z_2)_B\bar{\sigma}_B(z_3)\sigma_B(z_4)\rangle=\\
&=\dfrac{1}{2(z-w)^2}\left(\dfrac{z_{31}z_{42}}{z_{21}z_{41}z_{32}z_{43}}\right)^{1/8}\left[\sqrt{\dfrac{(z-z_1)(w-z_2)(z-z_3)(w-z_4)}{(w-z_1)(z-z_2)(w-z_3)(z-z_4)}}+(z \leftrightarrow w)\right]\,.
\end{split}
\end{equation}
Let us now compare \eqref{eq:jjssssB} to the sum of \eqref{eq:jjssss} over the array. Summing the first term gives simply
\begin{equation}
\dfrac{G_B(z_i)}{2(z-w)^2}\left[\sqrt{\dfrac{(z-z_1)(w-z_2)(z-z_3)(w-z_4)}{(w-z_1)(z-z_2)(w-z_3)(z-z_4)}}+(z \leftrightarrow w)\right]\,,
\end{equation}
where $G_B(z_i)=\langle\bar{\sigma}_B(z_1)\sigma_B(z_2)\bar{\sigma}_B(z_3)\sigma_B(z_4)\rangle$. The second term is proportional to
\begin{equation}
\partial_\eta\left[\dfrac{1}{\sqrt{K(\eta)}}\exp\left(\frac{i\delta^2}{8\pi}\tau(\eta)\right)\right]\,.
\end{equation}
Summing over $\delta=2\sqrt{2}\pi n$ we get $\partial_\eta \sqrt{\frac{2}{\pi}}=0$. Putting all together we notice that \eqref{eq:jjssssB} is recovered.\\
\noindent
Correlation functions involving excited twist fields, in particular $\langle\bar{\sigma}'_B(z_1)\sigma'_B(z_2)\bar{\sigma}_B(z_3)\sigma_B(z_4)\rangle$ and $\langle\bar{\sigma}'_B(z_1)\sigma_B(z_2)\bar{\sigma}'_B(z_3)\sigma_B(z_4)\rangle$ are computed in appendix \ref{app:jj}, and agree with the sum over the array of $\langle\bar{\sigma}'(z_1)\sigma'(z_2)\bar{\sigma}(z_3)\sigma(z_4)\rangle$ and $\langle\bar{\sigma}'(z_1)\sigma(z_2)\bar{\sigma}'(z_3)\sigma(z_4)\rangle$ respectively. The results of this section give further support to our claim that bosonized twist fields describe an array of Dirichlet sectors.

\subsection{Correlation functions with more than four bosonized twist fields}
The calculation of correlation functions becomes increasingly more complicated when the number of twist fields is more than four. Some results have been derived through the electrostatic analogy (see \cite{Frohlich} for $\mathbb{Z}_2$ twist fields and \cite{Abel:2003yx,Pesando:2012cx,Pesando} for generic angle twist fields); this procedure is however quite formal, since an explicit expression for the Green's function is known only in integral form. Furthermore, the generalization of the methods described in appendices \ref{app:4pt} and \ref{app:jj} is problematic.

In the bosonized case, however, it is still possible to compute correlation functions involving bosonized twist fields and, possibly, the current $\partial X$. For example the $2n$-point function of twist fields is given by
\begin{equation}\label{eq:2n_bos}
\langle\bar{\sigma}_B(z_1)\sigma_B(z_2)\dots\bar{\sigma}_B(z_{2n-1})\sigma_B(z_{2n})\rangle=\prod_{\substack{i>j \\ (i-j)\in2\mathbb{N}}} z_{ij}^{1/8}\prod_{\substack{i>j \\ (i-j)\in2\mathbb{N}+1}} z_{ij}^{-1/8}=\prod_{i>j} z_{ij}^{\frac{(-1)^{i-j}}{8}}\,.
\end{equation}
The correlator with two currents reads
\begin{equation}
\langle j(z)j(w)\bar{\sigma}_B(z_1)\dots\sigma_B(z_{2n})\rangle=\dfrac{G_B^{2n}(z_i)}{2(z-w)^2}\left(\prod_{i\text{ odd}}\sqrt{\dfrac{z-z_i}{w-z_i}}\prod_{i\text{ even}}\sqrt{\dfrac{w-z_i}{z-z_i}}+(z\leftrightarrow w)  \right)\,,
\end{equation}
where $G_B^{2n}(z_i)$ is the $2n$-point function \eqref{eq:2n_bos}. Correlators involving excited twist fields can also be considered; for example the correlation function of two excited and four normal twist fields is
\begin{equation}
\langle\bar{\sigma}'_B(z_1)\sigma'_B(z_2)\bar{\sigma}_B(z_3)\sigma_B(z_4)\bar{\sigma}_B(z_5)\sigma_B(z_6)\rangle=\dfrac{1}{2z_{21}^{9/8}}\left(\dfrac{z_{53}z_{64}}{z_{43}z_{63}z_{54}z_{65}}\right)^{1/8}\left(\dfrac{z_{31}z_{51}z_{42}z_{62}}{z_{32}z_{52}z_{41}z_{61}}\right)^{3/8}\,.
\end{equation}
Through the bosonization procedure one might easily see if a correlation function vanishes; this happens whenever the sum of all the exponents of operators $V_\alpha$ can not give zero. For example, a correlator with one excited and $2n-1$ normal twist fields is always zero:
\begin{equation}
\langle\bar{\sigma}'_B(z_1)\sigma_B(z_2)\dots\bar{\sigma}_B(z_{2n-1})\sigma_B(z_{2n})\rangle=0\,.
\end{equation}
The same is true for a correlator of $m$ excited and $2n-m$ normal twist fields, when $m$ is odd. Analogously, a correlator involving two excited twist fields vanishes if they are both conjugated (or both non-conjugated). For example
\begin{equation}
\langle\bar{\sigma}'_B(z_1)\sigma_B(z_2)\bar{\sigma}'_B(z_3)\dots\bar{\sigma}_B(z_{2n-1})\sigma_B(z_{2n})\rangle=0\,.
\end{equation}
Furthermore, every correlation function with an odd number of currents and $2n$ normal twist fields is zero:
\begin{equation}
\langle j(w_1)\dots j(w_{2m+1})\bar{\sigma}_B(z_1)\dots\sigma_B(z_{2n})\rangle=0\,.
\end{equation}
We have seen in appendices \ref{app:4pt} and \ref{app:jj} that these correlation functions (with $n=2$) are not vanishing for normal (non-bosonized) twist fields. The setup with the array of Dirichlet sectors is special, since it makes many correlation functions vanish.

\section{Ordering of boundary twist fields}\label{sec:ordering}
In the previous section, when computing correlation functions with twist fields on the boundary, we have always implicitly assumed a particular ordering of the twist fields. This is because a twist field connects a CFT to a different one (corresponding to different boundary conditions for the boson), and it must be followed by a conjugated twist field, in such a way that the correlation function is computed with respect to the vacuum of the original CFT. For concreteness, let us consider the two-point function of twist fields $\langle\bar{\sigma}(z)\sigma(w)\rangle$. The two twist fields connect the boundary conformal field theory of a boson with Dirichlet boundary condition (BCFT$_D$) to the boundary conformal field theory of a boson with Neumann boundary condition (BCFT$_N$); more precisely, reading the correlation function from left to right, $\bar{\sigma}$ connects BCFT$_D$ to BCFT$_N$ and $\sigma$ connects BCFT$_N$ to BCFT$_D$. To make things clear, we will indicate explicitly with $N$ or $D$ the vacuum of the reference CFT, which is the BCFT$_D$ in this case. Therefore, the correlation function has to be interpreted as
\begin{equation}
\langle\bar{\sigma}(z)\sigma(w)\rangle_D=\dfrac{\langle\unit\rangle_D}{(z-w)^{1/8}}=\dfrac{Z_D}{(z-w)^{1/8}}\,,
\end{equation}
where we have used the OPE \eqref{eq:ope_twist}, and $Z_D$ is the partition function in BCFT$_D$. If we now want to consider the opposite ordering, we have to consider the BCFT$_N$ as reference theory. Assuming that the OPE is
\begin{equation}
\sigma(z)\bar{\sigma}(w)=\dfrac{\alpha}{(z-w)^{1/8}}\,,
\end{equation}
we will have $\langle\sigma(z)\bar{\sigma}(w)\rangle_N=\alpha(z-w)^{-1/8}\langle\unit\rangle_N=\alpha(z-w)^{-1/8} Z_N$, where $Z_N$ is the partition function in BCFT$_N$. Since we are considering the twist fields inserted on the boundary of a disk (or, alternatively, on the compactified real line), cyclicity implies that $\langle\bar{\sigma}(z)\sigma(w)\rangle_D=\langle\sigma(w)\bar{\sigma}(z)\rangle_N$, from which we conclude that
\begin{equation}
\alpha=Z_D/Z_N\,.
\end{equation}
On the other hand, the number $\alpha$ can be computed using the four-point function. In fact
\begin{equation}
\lim_{z_2\rightarrow z_3}(z_2-z_3)^{1/8}\langle\bar{\sigma}(z_1)\sigma(z_2)\bar{\sigma}(z_3)\sigma(z_4)\rangle_D=\alpha\langle\bar{\sigma}(z_1)\sigma(z_4)\rangle_D=\dfrac{Z_D^2/Z_N}{(z_1-z_4)^{1/8}}\,.
\end{equation}
In the previous section we have chosen to normalize $Z_D=1$. The explicit form of the four-point function gives the result
\begin{equation}
\lim_{z_2\rightarrow z_3}(z_2-z_3)^{1/8}\langle\bar{\sigma}(z_1)\sigma(z_2)\bar{\sigma}(z_3)\sigma(z_4)\rangle_D=0\,,
\end{equation}
which means that the partition function $Z_N$ is divergent. However, $Z_N$ can be regularized, for example compactifying the boson on a circle of radius $R$, which would give $Z_N=1/R$ (see\cite{Erler}). 

The cyclical property we used for the two-point function generalizes to more complicated correlation functions of twist fields, provided that the appropriate reference CFT is taken into account. For example, for the four-point function,
\begin{equation}
\langle\bar{\sigma}(z_1)\sigma(z_2)\bar{\sigma}(z_3)\sigma(z_4)\rangle_D=\langle\sigma(z_2)\bar{\sigma}(z_3)\sigma(z_4)\bar{\sigma}(z_1)\rangle_N\,.
\end{equation}
With this rule, every correlation function with an even number of boundary twist fields, and with alternating $\sigma$'s and $\bar{\sigma}$'s has a precise and unambiguous meaning.

\section{Bulk twist fields and modular invariance}\label{sec:partition}
In appendix \ref{app:hyper} we review how the cuts created by the presence of twist fields have the effect of transforming the worldsheet into a higher genus Riemann surface. It is thus natural to think that the correlation function of twist fields, without any other operator, is associated to the partition function of a twisted boson on this surface. This was examined for example in \cite{Dijkgraaf} and \cite{Yang}. In order to connect to this result, we have to consider bulk twist fields, twisting both the chiral ($X$) and anti-chiral ($\bar{X}$) parts of the boson. The bulk twist fields are given by the product of a chiral and an anti-chiral twist fields. Effectively, a correlation function of bulk twist fields is given by the square of the correlation function of chiral twist fields.
An important observation that we have to make is that the Riemann surface is not sensible to which points the cuts are connecting and to which twist fields are conjugated and which are not. For concreteness, if we indicate (12)(34) the correlation function $\langle\bar{\sigma}(z_1)\sigma(z_2)\bar{\sigma}(z_3)\sigma(z_4)\rangle$, where the cuts are connecting $z_1$ to $z_2$ and $z_3$ to $z_4$, we see that the combination (34)(12), (21)(43) and (43)(21) describe the same situation. In total there are 6 independent ways of  partitioning the points $z_i$ in two non-ordered pairs, which correspond to different conformal ratios and, correspondingly, to different periods of the associated torus (see table \ref{tab:modular}).
\begin{table}[h!]
\begin{center}
  \begin{tabular}{ c | c | c }
    Partition & Conformal ratio & Period \\
    \hline
    (12)(34) & $\eta$ & $\tau$ \\ \hline
    (14)(32) & $1-\eta$ & -$\frac{1}{\tau}$ \\ \hline
    (12)(43) & $\frac{\eta}{\eta-1}$ & $\tau+1$ \\ \hline
    (13)(42) & $\frac{1}{1-\eta}$ & $-\frac{1}{\tau+1}$ \\ \hline
    (13)(24) & $\frac{1}{\eta}$ & $\frac{\tau}{1-\tau}$ \\ \hline
    (14)(23) & $\frac{\eta-1}{\eta}$ & $-\frac{1}{\tau}+1$ \\
    \hline
  \end{tabular}
  \caption{Partition of four points and modular transformations.}\label{tab:modular}
\end{center}
\end{table}
In order to recover the partition function on the torus, one should sum over all these independent partitions. Looking at the associated periods, we notice that the different partitions generate modular transformations on the period $\tau$. To be precise, the modular group can be generated by only two transformations:
\begin{equation}
\begin{split}
&S: \qquad \tau\rightarrow -\dfrac{1}{\tau}\,,\qquad\quad \eta\rightarrow 1-\eta\,,\\
&T: \qquad \tau\rightarrow \tau+1\,,\qquad\, \eta\rightarrow \dfrac{\eta}{\eta-1}\,.
\end{split}
\end{equation}
We refer to \cite{Dijkgraaf} for the proof that the sum
\begin{equation}\label{eq:PartitionFunction}
Z\propto \vert\langle\bar{\sigma}(z_1)\sigma(z_2)\bar{\sigma}(z_3)\sigma(z_4)\rangle\vert^2 +\text{permutations}
\end{equation}
is indeed the partition function of a twisted boson on a torus. Here we just notice that the result \eqref{eq:PartitionFunction} is manifestly modular invariant.

The same discussion can be done for bosonized twist fields; the resulting modular invariant partition function will be the one corresponding to a twisted boson on an orbifold of radius $R=\sqrt{2}$. This partition function and the one obtained by normal twist fields are related. The quantum part of the partition function (which depends only on the local property of twist fields) is the same, while the classical part, which depends on the topology of the surface, is different. In order to obtain the classical part, one has to sum over all the classical solutions in the different winding sectors around the circle (see e.g. \cite{Dijkgraaf}):
\begin{equation}
Z^{cl}(R)=\sum_{(p,\bar{p})}\exp\left[i\pi(p\cdot\tau\cdot p-\bar{p}\cdot\bar{\tau}\cdot\bar{p})\right]\,,
\end{equation}
where $p$ and $\bar{p}$ are the allowed momenta running through the loops of the hyperelliptic surface. It was noticed in \cite{Zamolodchikov} that when the radius of the compactification is exactly $\sqrt{2}$, the total partition function simplifies, and can be expressed in terms of correlation functions of operators of the form $:\exp(\alpha\phi(z)):$, where $\phi$ is a scalar field. The bosonisation introduced in this paper makes it clear that this scalar field is not the boson $X$, but it is the dual boson $\Omega$, and that the operators $:\exp(\alpha\phi(z)):$ are just our bosonized twist fields.

If one wants to insert the twist fields on the boundary, and interpret them as boundary changing operators, not all the partitions of table \ref{tab:modular} are allowed. As we discussed in section \ref{sec:ordering}, only the partitions (12)(34) and (14)(32) are well defined. This means that summing over the allowed partitions would give a result which is invariant only under the subgroup of the modular group generated by the $S$ transformation. This is consistent with the fact that, if the four twist fields are inserted on the real line, the associated period is purely imaginary, and a $T$ transformation would spoil this property.
 
\section{Application to string theory: bound state of D-branes}\label{sec:string}
Twist fields are useful in string theory in order to describe bound states of D-branes with different dimensions. Let us consider a bound state of a D$n$ and a D$m$ brane in bosonic string theory, with $n<m$. The coordinates $X^M$ of open string attached to the branes obey different boundary conditions. On the D$n$ brane, for example, the first coordinates $X^M$ ($M=0,\dots,n$) obey Neumann b.c. while the remaining $X^M$ ($M>n$) obey Dirichlet b.c. An open string stretching between the two different branes has mixed boundary conditions along the directions $X^M$ with ($n<M\leq m$); along these directions, the change in boundary conditions has to be taken into account, and twist fields must be inserted inside correlation functions.

\subsection{Boundary changing operators}
In the following we will focus on the bound state of a D$(-1)$ and a D$(n-1)$ brane; other bound states with a difference of dimensionality equal to $n$ can be related to this by a T-duality. A higher number of parallel D$(-1)$ and D$(n-1)$ branes can be also considered, but we will restrict ourselves to one for simplicity. Let us define the boundary changing operators $\Delta(z)$ (and $\bar{\Delta}$) as the product of the twist fields along the ``mixed''  directions, i.e.
\begin{equation}
\Delta(z)=\prod_{\mu=0}^{n-1}\sigma^\mu(z)\,, \qquad \bar{\Delta}(z)=\prod_{\mu=0}^{n-1}\bar{\sigma}^\mu(z)\,.
\end{equation}
Boundary changing operators are primaries of conformal dimension $n/16$, and satisfy
\begin{equation}
\bar{\Delta}(z)\Delta(w)=\dfrac{1}{(z-w)^{n/8}}\,.
\end{equation}
Correlation functions involving boundary changing operators can be derived from correlation functions with twist fields. When the difference of dimension equals some particular values, correlation functions can become quite simple. This is the case when $n$ is a multiple of four. We start considering co-dimension 8 and 16, and comment on the $n=4$ case at the end, due to its importance in superstring theory.

\subsection{D7-D(-1) system}
Let us now consider the difference of dimension to be 8, in particular the bound state of a D7 and a D(-1) brane. The four-point function of boundary changing operator has a simple form, namely
\begin{equation}
\langle\bar{\Delta}(z_1)\Delta(z_2)\bar{\Delta}(z_3)\Delta(z_4)\rangle=\left(\dfrac{z_{31}z_{42}}{z_{21}z_{41}z_{32}z_{43}}\right)\left(\dfrac{\pi}{2K(\eta)}\right)^4\,.
\end{equation}
The bosonized version of twist fields can also be used, but it describes a different setup. The periodicity properties of the boson $\Omega$ can be used for describing a set of D$(-1)$ branes positioned on a lattice with period $2\sqrt{2}\pi$. A pair of boundary changing operators connects the D7 brane to one of these D$(-1)$ branes; the four-point function can then depend on the positions of two different branes. If the difference of the two positions is given by the vector $\vec{\delta}$, the four-point function is
\begin{equation}
\langle\bar{\Delta}(z_1)\Delta(z_2)\bar{\Delta}(z_3)\Delta(z_4)\rangle=\left(\dfrac{z_{31}z_{42}}{z_{21}z_{41}z_{32}z_{43}}\right)\left(\dfrac{\pi}{2K(\eta)}\right)^4\exp\left(\dfrac{i|\vec{\delta}|^2\tau(\eta)}{8\pi}\right)\,.
\end{equation}
The position of every brane on the lattice can be described by a vector of four integer numbers $n^\mu$, i.e. $x^\mu=2\sqrt{2}\pi n^\mu$. The correlation function of bosonized twist fields is then given by the superposition of four-point function corresponding to single branes, the result being
\begin{equation}
\langle\bar{\Delta}_B(z_1)\Delta_B(z_2)\bar{\Delta}_B(z_3)\Delta_B(z_4)\rangle=\dfrac{z_{31}z_{42}}{z_{21}z_{41}z_{32}z_{43}}\,.
\end{equation}
Notice that in this case (co-dimension $n=8$) the boundary changing operator $\Delta$ has conformal dimension $1/2$. We want now to argue that, at least in the bosonized case, it behaves effectively as a fermion. From the eight bosons $\Omega^\mu$, we can construct the normalized boson $\Omega_{CM}$ as
\begin{equation}
\Omega_{CM}=\dfrac{1}{\sqrt{8}}\sum_{\mu=1}^{8}\Omega^\mu\,.
\end{equation}
Given this definition, the boundary changing operator can be written as
\begin{equation}
\Delta_B(z)=\exp\left(i\dfrac{\sqrt{2}}{4}\sum_{\mu=1}^{8}\Omega_i(z)\right)=e^{i\Omega_{CM}}(z)\,.
\end{equation}
We notice that this expression represents a complex fermion (in its bosonized representation).

\subsection{D15-D(-1) system}
Another notable situation we are considering is when the difference of dimension is 16. The four-point function of boundary changing operators is simply (allowing D(-1) branes at different positions)
\begin{equation}
\langle\bar{\Delta}(z_1)\Delta(z_2)\bar{\Delta}(z_3)\Delta(z_4)\rangle=\left(\dfrac{z_{31}z_{42}}{z_{21}z_{41}z_{32}z_{43}}\right)^2\left(\dfrac{\pi}{2K(\eta)}\right)^8\exp\left(\dfrac{i|\vec{\delta}|^2\tau(\eta)}{8\pi}\right)\,,
\end{equation}
while in the bosonized case the result is
\begin{equation}
\langle\bar{\Delta}_B(z_1)\Delta_B(z_2)\bar{\Delta}_B(z_3)\Delta_B(z_4)\rangle=\left(\dfrac{z_{31}z_{42}}{z_{21}z_{41}z_{32}z_{43}}\right)^2\,.
\end{equation}
The boundary changing operator has conformal dimension 1, and can be written (in the bosonized case), as
\begin{equation}
\Delta_B(z)=\exp\left(i\dfrac{\sqrt{2}}{4}\sum_{\mu=1}^{16}\Omega_i(z)\right)=e^{i\sqrt{2}\Omega_{CM}}(z)=:J^+_{CM}(z)\,,
\end{equation}
where $\Omega_{CM}=\sum\Omega^{\mu}/\sqrt{16}$ and $J^+_{CM}(z)$ is a generator of the current algebra described in section \ref{sec:bos_twist}. A natural question to ask is whether this dimension 1 operator can generate an exactly marginal deformation of the boundary conformal field theory.  We have to remember, however, that a twist field must always appear together with its conjugate
\begin{equation}
\bar{\Delta}_B(z)=e^{-i\sqrt{2}\Omega_{CM}}(z)=:J^-_{CM}(z)\,.
\end{equation}
This means that the deformation of the boundary CFT is given by 
\begin{equation}
\exp\left(\lambda^2\int J^+_{CM}(z)dz\int J^-_{CM}(w)dw\right)\,,
\end{equation}
where $\lambda$ is the modulus of the deformation. As discussed in \cite{Recknagel}, a set of dimension 1 boundary operators produces a marginal deformation only if these operators are mutually local, meaning that the OPE among them must not contain single poles. A similar result for bulk deformation states that a set of operators of the form $J_i(z)\bar{J_i}(\bar{z})$ generates an exactly marginal deformation of the theory if and only if these currents form an abelian subalgebra (see e.g. \cite{Chaudhuri,Forste}). In our case, however, we have
\begin{equation}
\bar{\Delta}_B(z)\Delta_B(w)=J^-_{CM}(z)J^+_{CM}(w)=\dfrac{1}{(z-w)^2}-\dfrac{i\sqrt{2}\partial\Omega_{CM}}{z-w}+\dots\,,
\end{equation}
which means that $\bar{\Delta}_B$ and $\Delta_B$ are not mutually local. Equivalently, $J^+_{CM}$ and $J^-_{CM}$ do not constitute a subalgebra of the $\mathfrak{su}(2)$ Ka\v{c}-Moody, since $[J^+_{CM},J^-_{CM}]\sim\partial\Omega_{CM}$. In conclusion, even if the boundary changing operator has conformal dimension 1, it does not generate an exactly marginal deformation of the bosonic conformal theory. Geometrically, the deformation generated by the twist field $\Delta_B(z)$ (which is the massless excitation of the $(-1,15)$ string) corresponds to blowing up the point-like D(-1) branes inside the D15 brane. We then conclude that this blowing up mode is not a modulus in the lattice. 

One may wonder if this obstruction is an artifact of compactification. Recalling the OPE (\ref{dope}) of the original twist field we see that a simple pole will be present whenever the compactification radius is a multiple of $\sqrt{2}$. So, we expect the obstruction to persist if this condition is met. A possible interpretation for the lifting of this modulus from string theory is that the the constituents of the array feel each other through the exchange of a massless primary.

\subsection{Superstring theory and correlation functions for the D3-D(-1) system}
In superstring theory the full boundary changing vertex operators contains also spin fields and ghosts. Due to picture changing, one also encounters the ``excited'' bosonic boundary changing operator, which consists of the product of one excited twist field and $n-1$ normal ones. More specifically we define
\begin{equation}
\tau^\mu=\sigma'^\mu(z)\prod_{\substack{\nu=0 \\ \nu\neq\mu}}^{n-1}\sigma^\nu(z)\,, \qquad \bar{\tau}^\mu=\bar{\sigma}'^\mu(z)\prod_{\substack{\nu=0 \\ \nu\neq\mu}}^{n-1}\bar{\sigma}^\nu(z)\,.
\end{equation}
Excited boundary changing operators are primaries of conformal dimension $n/16+1/2$; the operator product expansions can be easily derived from the ones defining the twist fields; for example
\begin{equation}
\begin{split}
&i\partial X^\mu(z)\Delta(w)=\dfrac{\tau^\mu(w)}{(z-w)^{1/2}}+\dots\,,\\
&i\partial X^\mu(z)\tau^\mu(w)=\dfrac{\Delta(w)}{2(z-w)^{3/2}}+\dfrac{2\ \partial\Delta(w)}{(z-w)^{1/2}}+ \dots\,,
\end{split}
\end{equation}
where we are not summing over the index $\mu$ in the second expression. Furthermore we have
\begin{equation}
\bar{\tau}^\mu(z)\tau^\nu(w)=\dfrac{\eta^{\mu\nu}}{2(z-w)^{n/8+1}}\,,
\end{equation}
where $\eta^{\mu\nu}$ is the metric of the target space.

Let us consider a system of the two branes with difference of dimension equal to 4; in particular we focus on a bound state of a D3 and (possibly many) D$(-1)$ branes. This system is relevant in superstring theory, in particular for its connection to gauge instantons (see e.g. \cite{Billo,Pesando:2011yd}). The calculation of four-point correlation functions is straightforward, and gives
\begin{equation}
\langle\bar{\Delta}(z_1)\Delta(z_2)\bar{\Delta}(z_3)\Delta(z_4)\rangle=\left(\dfrac{z_{31}z_{42}}{z_{21}z_{41}z_{32}z_{43}}\right)^{1/2}\left(\dfrac{\pi}{2K(\eta)}\right)^2\exp\left(\dfrac{i\vec{\delta}^2\tau(\eta)}{8\pi}\right),
\end{equation}
\begin{equation}
\begin{split}
\langle\bar{\tau}^\mu(z_1)\tau^\nu(z_2)\bar{\Delta}(z_3)\Delta(z_4)\rangle=&\\
\dfrac{\eta^{\mu\nu}}{z_{21}^{3/2}z_{43}^{1/2}}\left(\dfrac{\pi}{2K(\eta)}\right)^2&\dfrac{1}{1-\eta}\left(\dfrac{E(\eta)}{2K(\eta)}-\dfrac{\vec{\delta}^2}{16K(\eta)^2}\right)\exp\left(\dfrac{i\vec{\delta}^2\tau(\eta)}{8\pi}\right)\,,\\
\langle\bar{\tau}^\mu(z_1)\Delta(z_2)\bar{\tau}^\nu(z_3)\Delta(z_4)\rangle=&\\
=\dfrac{\eta^{\mu\nu}z_{42}^{1/2}}{z_{31}^{1/2}z_{43}z_{21}}\left(\dfrac{\pi}{2K(\eta)}\right)^2&\dfrac{1}{1-\eta}\left(\dfrac{1-\eta}{2}-\dfrac{E(\eta)}{2K(\eta)}+\dfrac{\vec{\delta}^2}{16K(\eta)^2}\right)\exp\left(\dfrac{i\vec{\delta}^2\tau(\eta)}{8\pi}\right)\,,
\end{split}
\end{equation}
where $K(\eta)$ and $E(\eta)$ are the complete elliptic integrals of the first and second kind respectively. Correlation functions of bosonized twist fields are then given by the superposition of four-point functions corresponding to single branes, the results being
\begin{equation}
\begin{split}
&\langle\bar{\Delta}_B(z_1)\Delta_B(z_2)\bar{\Delta}_B(z_3)\Delta_B(z_4)\rangle=\left(\dfrac{z_{31}z_{42}}{z_{21}z_{41}z_{32}z_{43}}\right)^{1/2}\,,\\
&\langle\bar{\tau}^\mu_B(z_1)\tau^\nu_B(z_2)\bar{\Delta}_B(z_3)\Delta_B(z_4)\rangle=\dfrac{\eta^{\mu\nu}}{2z_{21}^{3/2}z_{43}^{1/2}}\,,\\
&\langle\bar{\tau}^\mu_B(z_1)\Delta_B(z_2)\bar{\tau}^\nu_B(z_3)\Delta_B(z_4)\rangle=0\,.
\end{split}
\end{equation}
These correlation functions can also be derived in a straightforward way by expressing the boundary changing operators in the $\Omega$ picture.

\section{Conclusions}
The primary motivation of this project is to explore the moduli space of bound states of D-branes, both in bosonic and superstring theory. The key ingredient for this is the OPE of twist fields and various correlation functions containing a higher number of twist fields. Therefore, the purpose of this note  is to collect and extend results on correlation functions containing an arbitrary number of $\mathbb{Z}_2$ boundary changing operators for a free boson. These are  the correlators that arise primarily when considering bound states of D-branes in string theory. Of course, in superstring theory, the relevant correlators contain other fields such as spin fields, ghost fields etc., but the complication in the calculation of such correlators resides mainly in the bosonic twist fields. In particular,  higher twist field insertions become important, for instance, when considering bound states of finite size D-branes, which in turn are related to the instanton moduli space in the field theory limit. On another front they enter crucially when considering effective actions \cite{Maccaferri} and classical solutions in string field theory \cite{Erler}, which was one motivation for the present work. While adding some explicit results to the list of known correlations functions containing $\mathbb{Z}_2$ twist fields, the key result of the present paper is the bosonization of bosonic twist fields which we argued to describe an array of D-brane bound states based on their relation to orbifold theories. As an application we were able to show that the modulus corresponding to the size of the D(-1) branes bound to a D15 brane is obstructed, since this deformation is equivalent to a marginal deformation of two non-commuting chiral currents. An interesting question is to extend this analysis to the bound state of BPS branes in superstring theory. We will return to this question in \cite{ToCome}.

 \bigskip
\noindent{\bf Acknowledgements:}\\
\smallskip
We would like to thank Igor Pesando for many constructive comments on the draft as well as Sebastian Konopka, Carlo Maccaferri, Alberto Merlano and Tom\'{a}\v{s} Proch\'{a}zka for inspiring  discussions. This work  was supported by the DFG Transregional Collaborative Research Centre TRR 33 and the DFG cluster of  excellence "Origin and Structure of the Universe".

\newpage
\appendix
\section{Electrostatic analogy}\label{app:electrostatics}
The results of section \ref{sec:corr_2_twsit} (for the flat space case) can be interpreted in the language of electrostatics in two dimensions \cite{Corrigan:1975sn}. In absence of twist fields the Green's function for the Laplace operator is $G(z,w)=\log(z-w)$, satisfying $\bigtriangleup_zG(z,w)=2\pi\delta^{(2)}(z-w)$. The OPE's among the fields can be written in terms of the Green's function and its derivatives, for example
\begin{equation}\label{eq:various_OPE}
\begin{split}
&X(z)X(w)\sim -G(z,w)\sim -\log(z-w)\,,\\
&j(z)j(w)\sim \partial_z\partial_w G(z,w)\sim \dfrac{1}{(z-w)^2}\,,\\
&j(z)\widetilde{V}_\alpha(w)\sim\alpha\,\partial_zG(z,w)\, \widetilde{V}_\alpha(w)\sim\dfrac{\alpha}{(z-w)}\widetilde{V}_\alpha(w)\,.
\end{split}
\end{equation}
The correlation function of many primaries of the form $\widetilde{V}_\alpha$ is then given by
\begin{equation}\label{eq:corr_primaries}
\langle\widetilde{V}_{\alpha_1}(z_1)\dots\widetilde{V}_{\alpha_n}(z_n)\rangle=\exp\left(\sum_{i<j}\alpha_i\alpha_jG(z_i,z_j)\right)\delta\left(\sum_{i=1}^n \alpha_i\right)=\prod_{i<j}(z_{ij})^{\alpha_i \alpha_j}\,\,\delta\left(\sum_{i=1}^n \alpha_i\right)\,.
\end{equation}
The delta function is a consequence of the integration over the zero modes; furthermore, there is no contribution proportional to $\alpha_i^2$, since it is always possible to take a flat metric in a large region containing all the vertex operators (see \cite{Polchinski_book} for a discussion about that). 

The Green's function in the presence of two twist fields can be derived using the method of image charges, as in \cite{Hashimoto}. The result, for Dirichlet and Neumann boundary conditions on the positive and negative real axis respectively, is
\begin{equation}
G(z,w)=\log\left(\dfrac{1-\sqrt{\frac{z}{w}}}{1+\sqrt{\frac{z}{w}}}\right)=\log\left(\dfrac{\sqrt{w}-\sqrt{z}}{\sqrt{w}+\sqrt{z}}\right)\,.
\end{equation}
Analogously to \eqref{eq:various_OPE} we derive
\begin{equation}\label{eq:green_2twist}
\begin{split}
&X(z)X(w)\sim -G(z,w)\sim -\log\left(\dfrac{1-\sqrt{\frac{z}{w}}}{1+\sqrt{\frac{z}{w}}}\right)\,,\\
&j(z)j(w)\sim \partial_z\partial_w G(z,w)\sim \dfrac{1}{2(z-w)^2}\left(\sqrt{\dfrac{z}{w}}+\sqrt{\dfrac{w}{z}}\right)\,,\\
&j(z)\psi_\alpha(w)\sim\alpha\,\partial_zG(z,w) \,\psi_\alpha(w)\sim\dfrac{\alpha}{(z-w)}\sqrt{\dfrac{w}{z}}\psi_\alpha(w)\,.
\end{split}
\end{equation}
The correlation function of many primaries $\psi_\alpha$ is slightly more complicated. As clarified in \cite{Frohlich}, the zero mode is absent but there is an extra contribution of the form
\begin{equation}
\exp\left(\sum_{i=1}^{n}\dfrac{\alpha_i^2}{2}\,S_0(z_i)\right)\,.
\end{equation}
It can be interpreted as a renormalized electrostatic self-energy and it takes care of the normal ordering discussed in section \ref{subsec:normal_ordering} (cfr. \cite{Mukhopadhyay}). $S_0$ is defined in general by
\begin{equation}
G(z,w)=\log(w-z)+S_0(z)+\mathcal{O}(w-z)\,.
\end{equation}
In the case of two twist fields at $0$ and $\infty$, \eqref{eq:green_2twist} gives $S_0(z)=\log(\frac{1}{4z})$, from which formulas like \eqref{eq:ssexp} and \eqref{eq:psipsi2twists} follow.

\section{Twist fields and hyperelliptic surfaces}\label{app:hyper}
Let us consider $2n$ twist fields at positions $z_i$ on the real line. We assume that the fields $\sigma$ are at position $z_i$ with $i$ even, and the fields $\bar{\sigma}$ correspond to odd $i$. The current $j$ has Neumann boundary conditions on the intervals $[z_{2i-1},z_{2i}]$, and Dirichlet boundary conditions on the intervals $[z_{2i},z_{2i+1}]$. The real line is to be considered as compactified, therefore there are Dirichlet boundary conditions also on the interval $[z_{2n},z_1]$, containing the point at infinity. The complex plane (described by a coordinate $z$) has cuts along the real axis, in correspondence to the intervals with Neumann boundary conditions. The associated hyperelliptic surface, which has genus $g=n-1$, is described by the equation
\begin{equation}\label{eq:hyper}
w^2=P(z):=\prod_{i=1}^{2n}(z-z_i)\,.
\end{equation}
Let us define some useful quantities; first of all we consider a canonical homology class $\{A_k,B_k\}$, where $A_k$ and $B_k$ are the A and B cycles of the hyperelliptic surface. In the description on the complex plane, these cycles surround two neighbouring ramification points. There is a basis for holomorphic 1-forms on this hyperelliptic surface, given by
\begin{equation}
\omega_k=\dfrac{z^{k-1}\,\mathrm{d}z}{\sqrt{P(z)}}\qquad \text{for} \quad k=1,\dots,g\,.
\end{equation}
Denoting with $A_k$ the A-cycle surrounding the two ramification points $z_{2k-1}$ and $z_{2k}$, the period of the 1-form $\omega_l$ along $A_k$ is defined as
\begin{equation}
\Omega_{kl}=\oint_{A_k}\omega_l=\oint_{A_k}\dfrac{z^{l-1}\,\mathrm{d}z}{\sqrt{P(z)}}\,.
\end{equation}
There is also a dual basis for holomorphic 1-forms $\zeta_l$, satisfying
\begin{equation}
\oint_{A_k}\zeta_l=\delta_{kl}\,.
\end{equation}
The period matrix of the hyperelliptic surface is defined in the following way:
\begin{equation}
\tau_{kl}=\oint_{B_k}\zeta_l\,,
\end{equation}
where $B_k$ is the B-cycle surrounding the two ramification points $z_{2k}$ and $z_{2k+1}.$

\subsection*{Four twist fields and the associated torus}
\begin{figure}[h!]
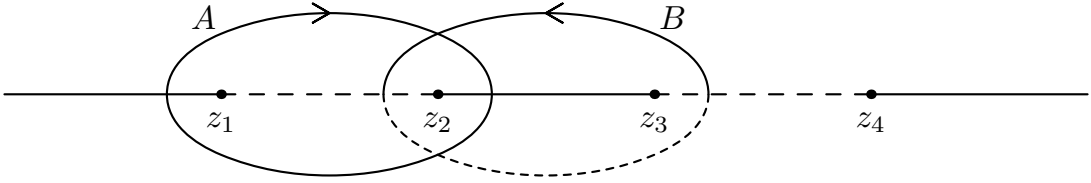

\vspace*{.8cm}
\centering
$\convertMPtoPDF{torus_complex.1}{.9}{.9}$
\caption{A and B cycles for a complex plane with 4 twist fields insertions.}
\label{fig:torus_complex}
\end{figure}

\begin{figure}[h!]
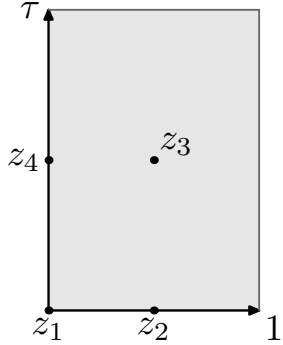

\centering
$\convertMPtoPDF{torus_flat.1}{1.}{1.}$
\caption{Fundamental domain of a torus in uniformized coordinates.}
\label{fig:torus_uniformized}
\end{figure}

When we have only four twist fields, the genus of the surface is $g=1$. This means that we are dealing with a torus, whose A and B cycles are shown in figure \ref{fig:torus_complex}.
\\
We have only one holomorphic 1-form
\begin{equation}
\omega=\dfrac{\mathrm{d}z}{\sqrt{P(z)}}\,,
\end{equation}
and its dual $\zeta$ given by
\begin{equation}
\zeta=\dfrac{\omega}{\Omega}=\left.\omega\middle/\oint_A\omega\right.\,.
\end{equation}
Therefore the period $\tau$ is simply
\begin{equation}
\left.\tau=\oint_B\omega\middle/ \oint_A\omega\right.\,.
\end{equation}
It can be useful to distinguish the two periods of the torus as $\tau_1=\oint_A\omega$ and $\tau_2=\oint_B\omega$, $\tau$ being the ratio of the two. Notice that, given the definition \eqref{eq:hyper} and assuming that the twist fields are inserted on the real line, the quantity $\tau_2$ is real, while $\Omega=\tau_1$ and $\tau$ are purely imaginary. Introducing the conformal cross ratio $\eta=z_{43}z_{21}/(z_{42}z_{31})$, where $z_{ij}=z_i-z_j$, the period can be written as
\begin{equation}
\tau=i\dfrac{K(1-\eta)}{K(\eta)}\,,
\end{equation}
where $K$ is the complete elliptic integral of the first kind. This relation can be inverted using Jacobi theta functions, namely
\begin{equation}
\eta=\left(\dfrac{\vartheta_2(0;\tau))}{\vartheta_3(0;\tau)}\right)^4\,.
\end{equation}
A positive, purely imaginary $\tau$ corresponds to $0<\eta<1$; the modular transformation $\tau\rightarrow -1/\tau$ corresponds to the map $\eta\rightarrow 1-\eta$. 
We will also use the so-called \textit{uniformized coordinates}, defined by
\begin{equation}
x(z)=\dfrac{1}{\Omega}\int_{z_1}^z\omega=\dfrac{1}{\Omega}\int_{z_1}^z\dfrac{dw}{\sqrt{P(w)}}\,. 
\end{equation}
In these coordinates the torus is flat, and we identify points on the complex plane via $x\equiv x+m +n\tau$, where $m,n\in\mathbb{Z}$. The four points $z_1$, $z_2$, $z_3$ and $z_4$ are mapped to $0$, $\tau/2$, $(\tau+1)/2$ and $1/2$ respectively. The torus can thus be described as the quotient
\begin{equation}
\mathbb{T}_2=\dfrac{\mathbb{C}}{\mathbb{Z}+\tau\mathbb{Z}}\,.
\end{equation}
The fundamental domain is shown in figure \ref{fig:torus_uniformized}.

\section{Four-point function of twist fields}\label{app:4pt}
In this appendix we follow the procedure of \cite{Zamolodchikov} in order to compute the four-point function of twist fields
\begin{equation}
G(z_i)=\langle\bar{\sigma}(z_1)\sigma(z_2)\bar{\sigma}(z_3)\sigma(z_4)\rangle\,.
\end{equation}
As in appendix \ref{app:hyper}, we have two Dirichlet intervals [$z_4,z_1$] (which includes the point at infinity) and [$z_2,z_3$]. We consider the closed cycle $C$, that encircles the point $z_1$ and $z_2$. We assume furthermore that the cycle is symmetric with respect to the real axis. We have that
\begin{equation}
\oint_C dz\,\mathsf{j}(z)= \int_{C^{>}}dz\,j(z)-\int_{C_{<}}dz\,\bar{j}(z)=\int_{C^{>}}dz\,i(\partial+\bar{\partial})X(z,\bar{z})= i\delta X_0\,,
\end{equation}
where $\delta X_0$ is the difference between the zero modes of $X(z,\bar{z})$ on the two Dirichlet intervals. Consider now a new correlation function
\begin{equation}
\Gamma(w,z_i)=\langle \mathsf{j}(w)\bar{\sigma}(z_1)\sigma(z_2)\bar{\sigma}(z_3)\sigma(z_4)\rangle\,.
\end{equation}
Integrating around the circle $C$ we get the so-called block condition
\begin{equation}
\oint \dfrac{dw}{2\pi i}\, \Gamma(w,z_i)=pG(z_i)\,,
\end{equation}
where $p=\delta X_0/2\pi$. Considering the OPE defining the twist fields, we can use the following Ansatz for $\Gamma$:
\begin{equation}\label{eq:Ansatz1}
\Gamma(w,z_i)=[(w-z_1)(w-z_2)(w-z_3)(w-z_4)]^{-1/2}A(z_i)\,,
\end{equation}
where $A(z_i)$ does not depend on $w$. Performing now the limit $w\rightarrow z_2$, and using again the OPE, we find
\begin{equation}
\lim_{z\rightarrow z_2}\Gamma(w,z_i)=\dfrac{1}{(w-z_2)^{1/2}}G^{(2)}(z_i)+\dots\,,
\end{equation}
where $G^{(2)}(z_i)=\langle\bar{\sigma}(z_1)\sigma'(z_2)\bar{\sigma}(z_3)\sigma(z_4)\rangle$ and $\dots$ represent terms of order $(z-w)^{1/2}$. On the other hand \eqref{eq:Ansatz1} implies 
\begin{equation}
\lim_{z\rightarrow z_2}\Gamma(w,z_i)=\dfrac{1}{(w-z_2)^{1/2}}\dfrac{A(z_i)}{\sqrt{z_{21}z_{32}z_{42}}}+\dots\,.
\end{equation}
Comparing the two equations gives $A(z_i)=\sqrt{z_{21}z_{32}z_{42}}\,G^{(2)}(z_i)$. Consider now another correlation function, namely
\begin{equation}
\Gamma^{(2)}(w,z_i)=\langle \mathsf{j}(w)\bar{\sigma}(z_1)\sigma'(z_2)\bar{\sigma}(z_3)\sigma(z_4)\rangle\,.
\end{equation}
Integrating over $w$ around the cycle $C$ we obtain another block condition, that reads
\begin{equation}
\oint \dfrac{dw}{2\pi i}\, \Gamma^{(2)}(w,z_i)=pG^{(2)}(z_i)\,.
\end{equation}
Considering now the local properties when $w$ approaches the insertion points $z_i$, the proper Ansatz for $\Gamma^{(2)}$ is
\begin{equation}\label{eq:Ansatz2}
\Gamma^{(2)}(w,z_i)=[(w-z_1)(w-z_2)(w-z_3)(w-z_4)]^{-1/2}\left(\dfrac{B(z_i)}{w-z_2}+C(z_i)\right)\,.
\end{equation}
Expanding this for $w\rightarrow z_2$ we find
\begin{equation}
\lim_{z\rightarrow z_2}\Gamma^{(2)}(w,z_i)=\dfrac{1}{\sqrt{(w-z_2)z_{21}z_{32}z_{42}}}\left(\dfrac{B(z_i)}{w-z_2}+C(z_i)-\dfrac{1}{2}B(z_i)\left(\frac{1}{z_{21}}+\frac{1}{z_{23}}+\frac{1}{z_{24}}\right)\right)+\dots
\end{equation}
On the other hand, the OPE implies that
\begin{equation}
\lim_{z\rightarrow z_2}\Gamma^{(2)}(w,z_i)=\dfrac{1}{2(w-z_2)^{3/2}}G(z_i)+\dfrac{2}{(w-z_2)^{1/2}}\partial_{z_2}G(z_i)+\dots
\end{equation}
Comparing the last two equations we find closed expression for $B(z_i)$ and $C(z_i)$:
\begin{equation}
\begin{split}
&B(z_i)=\frac{1}{2}\sqrt{z_{21}z_{32}z_{42}}\,G(z_i)\,,\\
&C(z_i)=\sqrt{z_{21}z_{32}z_{42}}\left(\dfrac{1}{4}\left(\frac{1}{z_{21}}+\frac{1}{z_{23}}+\frac{1}{z_{24}}\right)+2\dfrac{\partial}{\partial x_2}\right)G(z_i)\,.
\end{split}
\end{equation}
Finally we use the relation
\begin{equation}
\mathcal{K}(z_i)=\oint_C dw [(w-z_1)(w-z_2)(w-z_3)(w-z_4)]^{-1/2}=\dfrac{4i}{\sqrt{z_{31}z_{42}}}K(\eta)\,,
\end{equation}
where $\eta=z_{43}z_{21}/(z_{42}z_{31})$, and $K(\eta)$ is the complete elliptic integral of the first kind. Using this we can rewrite the two block conditions as
\begin{equation}
\begin{split}
&A(z_i)\mathcal{K}(z_i)=2\pi i p G(z_i)\,,\\
&\left(C(z_i)+2B(z_i)\frac{\partial}{\partial x_2}\right)\mathcal{K}(z_i)=2\pi i p G^{(2)}(z_i)\,.
\end{split}
\end{equation}
Inserting the relations we found for $A$, $B$ and $C$ we finally find a differential equation for the original correlation function:
\begin{equation}
\mathcal{K}^{3/2}(z_i)\dfrac{\partial}{\partial x_2}\left[(z_{21}z_{32}z_{42})^{1/8}\mathcal{K}^{1/2}(z_i)G(z_i)\right]=-2\pi^2 p^2(z_{21}z_{32}z_{42})^{-7/8}G(z_i)\,,
\end{equation}
whose solution is
\begin{equation}
G(z_i)\propto\left(\dfrac{z_{31}z_{42}}{z_{21}z_{41}z_{32}z_{43}}\right)^{1/8}\dfrac{1}{\sqrt{K(\eta)}}\exp\left(\frac{i(\delta X_0)^2}{8\pi}\tau(\eta)\right)\,.
\end{equation}
Here $\tau(\eta)$ is given by $\tau(\eta)=iK(1-\eta)/K(\eta)$. The overall normalization factor can be fixed using the OPE of twist fields. Knowing that $\bar{\sigma}(z)\sigma(w)\sim (z-w)^{-1/8}+\dots$, we have to require that
\begin{equation}
\lim_{z_1\rightarrow z_2}G(z_i)(z_1-z_2)^{1/8}=(z_3-z_4)^{-1/8}.
\end{equation}
This fixes the overall factor to be $\sqrt{\dfrac{\pi}{2}}$. We summarize here the result for the four-point function of twist fields and the other correlators introduced for the derivation:
\begin{equation}
\begin{split}
G(z_i)&=\langle\bar{\sigma}(z_1)\sigma(z_2)\bar{\sigma}(z_3)\sigma(z_4)\rangle=\left(\dfrac{z_{31}z_{42}}{z_{21}z_{41}z_{32}z_{43}}\right)^{1/8}\sqrt{\dfrac{\pi}{2K(\eta)}}\exp\left(\frac{i(\delta X_0)^2}{8\pi}\tau(\eta)\right)\,,\\
\Gamma(w,z_i)&=\langle \mathsf{j}(w)\bar{\sigma}(z_1)\sigma(z_2)\bar{\sigma}(z_3)\sigma(z_4)\rangle=\\
&=\dfrac{1}{4}\sqrt{\dfrac{\pi}{2P(w)}}\dfrac{(z_{31}z_{42})^{5/8}}{(z_{21}z_{41}z_{32}z_{43})^{1/8}}\dfrac{\delta X_0}{K(\eta)^{3/2}}\exp\left(\frac{i(\delta X_0)^2}{8\pi}\tau(\eta)\right)\,,\\
G^{(2)}(z_i)&=\langle\bar{\sigma}(z_1)\sigma'(z_2)\bar{\sigma}(z_3)\sigma(z_4)\rangle=\\
&=\dfrac{1}{4}\sqrt{\dfrac{\pi}{2}}\left(\dfrac{z_{31}}{z_{21}z_{32}}\right)^{5/8}\left(\dfrac{z_{42}}{z_{41}z_{43}}\right)^{1/8}\dfrac{\delta X_0}{K(\eta)^{3/2}}\exp\left(\frac{i(\delta X_0)^2}{8\pi}\tau(\eta)\right)\,,\\
\Gamma^{(2)}(w,z_i)&=\langle \mathsf{j}(w)\bar{\sigma}(z_1)\sigma'(z_2)\bar{\sigma}(z_3)\sigma(z_4)\rangle=\\
&=\sqrt{\dfrac{\pi}{2P(w)K(\eta)}}z_{31}^{9/8}\left(\dfrac{z_{42}}{z_{21}z_{32}}\right)^{5/8}\left(\dfrac{1}{z_{41}z_{43}}\right)^{1/8}\delta X_0\cdot\\
&\qquad\qquad\cdot\left(\dfrac{w-z_1}{2(w-z_2)}\dfrac{z_{32}}{z_{31}}+\dfrac{E(\eta)}{2K(\eta)}-\dfrac{\delta^2}{16K(\eta)^2}\right)\exp\left(\frac{i(\delta X_0)^2}{8\pi}\tau(\eta)\right)\,.
\end{split}
\end{equation}
Here $P(w)$ indicates the product $P(w)=(w-z_1)(w-z_2)(w-z_3)(w-z_4)$. Notice that the three correlation functions $\Gamma(w,z_i)$, $G^{(2)}(z_i)$ and $\Gamma^{(2)}(w,z_i)$ are proportional to the difference $\delta X_0$; therefore, when summed over the array, they give vanishing results. This means that the bosonized version of these correlation functions are zero, as one could derive by direct calculation in the $\Omega$ picture.

\section{Correlation function with four twist fields and two currents}\label{app:jj}
In this appendix we consider the Green's function in the presence of four twist fields. In particular, following \cite{Dixon} and \cite{Gava:1997jt}, we compute (assuming that Im$z>0$ and Im$w>0$)
\begin{equation}
g(z,w,z_i)=\partial_z\partial_w G(z,w)=\dfrac{\langle j(z)j(w)\bar{\sigma}(z_1)\sigma(z_2)\bar{\sigma}(z_3)\sigma(z_4)\rangle}{\langle\bar{\sigma}(z_1)\sigma(z_2)\bar{\sigma}(z_3)\sigma(z_4)\rangle}\,.
\end{equation}
Taking in consideration the OPE among $j$ and the twist fields, we can make an Ansatz for $g$, which reads
\begin{equation}\label{eq:ansatz_jj}
g(z,w,z_i)=\dfrac{1}{2(z-w)^2}\left[\sqrt{\dfrac{(z-z_1)(w-z_2)(z-z_3)(w-z_4)}{(w-z_1)(z-z_2)(w-z_3)(z-z_4)}}+(z \leftrightarrow w)\right]+\dfrac{A(z_i)}{\sqrt{P(z)P(w)}}\,,
\end{equation}
where $P(z)=(z-z_1)(z-z_2)(z-z_3)(z-z_4)$. We now use the definition of energy-momentum tensor
\begin{equation}
T(z)=\dfrac{1}{2}N(jj)(z)=\dfrac{1}{2}\left(\lim_{w\rightarrow z}j(z)j(w)-\dfrac{1}{(z-w)^2}\right)\,;
\end{equation}
this implies that
\begin{equation}
\dfrac{1}{2}\left(\lim_{w\rightarrow z}j(z)j(w)-\dfrac{1}{(z-w)^2}\right)=\dfrac{\langle T(z)\bar{\sigma}(z_1)\sigma(z_2)\bar{\sigma}(z_3)\sigma(z_4)\rangle}{G(z_i)}\,.
\end{equation}
The direct calculation gives
\begin{equation}\label{eq:T4sigma}
\dfrac{\langle T(z)\bar{\sigma}(z_1)\sigma(z_2)\bar{\sigma}(z_3)\sigma(z_4)\rangle}{G(z_i)}=\dfrac{1}{2}\dfrac{A(z_i)}{P(z_i)}+\dfrac{1}{16}\left(\dfrac{1}{z-z_1}-\dfrac{1}{z-z_2}+\dfrac{1}{z-z_3}-\dfrac{1}{z-z_4}\right)^2\,.
\end{equation}
We can now use the OPE of $T$ with $\sigma(z_2)$, in order to get the condition
\begin{equation}
\lim_{z\rightarrow z_2}\langle T(z)\bar{\sigma}(z_1)\sigma(z_2)\bar{\sigma}(z_3)\sigma(z_4)\rangle=\dfrac{G(z_i)}{16(z-z_2)^2}+\dfrac{\partial_{z_2}G(z_i)}{z-z_2}\,.
\end{equation}
Performing the limit on \eqref{eq:T4sigma} gives the equation
\begin{equation}
\partial_{z_2} \log(G(z_i))=\dfrac{A(z_i)}{2z_{21}z_{23}z_{24}}-\dfrac{1}{8}\left(\dfrac{1}{z_{21}}+\dfrac{1}{z_{23}}-\dfrac{1}{z_{24}}\right)\,,
\end{equation}
from which we find the function $A(x)$:
\begin{equation}
A(z_i)=2z_{21}z_{23}z_{24}\dfrac{\partial}{\partial_{z_2}}\left[\log\left((z_{21}z_{23}/z_{24})^{1/8}G(z_i)\right)\right]=z_{42}z_{31}\left(\dfrac{1-\eta}{2}-\dfrac{E(\eta)}{2K(\eta)}+\dfrac{\delta^2}{16K(\eta)^2}\right)\,,
\end{equation}
where $\delta=\delta X_0$ and $K(x)$ and $E(x)$ are the complete elliptic integrals of the first and second kind respectively. Analogous equations can be found for $z_1$, $z_3$ and $z_4$. We can finally join the equations together, using the property (for any function $f(z_1,z_2,z_3,z_4)$)
\begin{equation}
\eta(1-\eta)\partial_\eta f(z_i)=\dfrac{1}{z_{42}z_{31}}\left(z_{12}z_{13}z_{14}\partial_{z_1}+z_{21}z_{23}z_{24}\partial_{z_2}+z_{31}z_{32}z_{34}\partial_{z_3}+z_{41}z_{42}z_{43}\partial_{z_4}\right)f(z_i)\,.
\end{equation}
The final compact expression for $A(z_i)$ is 
\begin{equation}
A(z_i)=2z_{42}z_{31}\eta(1-\eta)\partial_\eta\log\left[\dfrac{1}{\sqrt{K(\eta)}}\exp\left(\frac{i\delta^2}{8\pi}\tau(\eta)\right)\right]\,,
\end{equation}
which was found in \cite{Dixon}, in the case $\delta=0$. Multiplying \eqref{eq:ansatz_jj} by $G(z_i)$ we find the correlator
\begin{equation}
\begin{split}
\langle j(z)j(w)\bar{\sigma}&(z_1)\sigma(z_2)\bar{\sigma}(z_3)\sigma(z_4)\rangle=\\
&=\dfrac{G(z_i)}{2(z-w)^2}\left[\sqrt{\dfrac{(z-z_1)(w-z_2)(z-z_3)(w-z_4)}{(w-z_1)(z-z_2)(w-z_3)(z-z_4)}}+(z \leftrightarrow w)\right]+\\
&+\dfrac{\sqrt{2\pi}}{\sqrt{P(z)P(w)}}\left(\dfrac{z_{31}z_{42}}{z_{21}z_{41}z_{32}z_{43}}\right)^{-7/8}\partial_\eta\left[\dfrac{1}{\sqrt{K(\eta)}}\exp\left(\frac{i\delta^2}{8\pi}\tau(\eta)\right)\right]\,.
\end{split}
\end{equation}
Taking appropriate limits one can derive correlation functions involving excited twist fields. For example, we can recover the correlator $\langle j(w)\bar{\sigma}(z_1)\sigma'(z_2)\bar{\sigma}(z_3)\sigma(z_4)\rangle$, which was already computed in appendix \ref{app:4pt}. Considering the limit when both of the currents collide with a twist field we find correlators with two excited twist fields. For example
\begin{equation}\label{eq:ttss}
\begin{split}
\langle\bar{\sigma}'(z_1)\sigma'(z_2)&\bar{\sigma}(z_3)\sigma(z_4)\rangle=\\
&=\dfrac{1}{z_{21}^{9/8}z_{43}^{1/8}}\left(\dfrac{z_{31}z_{42}}{z_{41}z_{32}}\right)^{5/8}\left(\dfrac{E(\eta)}{2K(\eta)}-\dfrac{\delta^2}{16K^2(\eta)}\right)\sqrt{\dfrac{\pi}{2K(\eta)}}\exp\left(\frac{i\delta^2}{8\pi}\tau(\eta)\right)\,.
\end{split}
\end{equation}
 When the two excited twist fields are not adjacent we get
\begin{equation}\label{eq:tsts}
\begin{split}
\langle\bar{\sigma}'(z_1)\sigma&(z_2)\bar{\sigma}'(z_3)\sigma(z_4)\rangle=\\
&=\dfrac{z_{31}^{1/8}z_{42}^{9/8}}{(z_{43}z_{41}z_{32}z_{21})^{5/8}}\left(\dfrac{1-\eta}{2}-\dfrac{E(\eta)}{2K(\eta)}+\dfrac{\delta^2}{16K^2(\eta)}\right)\sqrt{\dfrac{\pi}{2K(\eta)}}\exp\left(\frac{i\delta^2}{8\pi}\tau(\eta)\right)\,.
\end{split}
\end{equation}
Correlation functions involving excited twist fields are easily computed summing \eqref{eq:ttss} and \eqref{eq:tsts} over the array of Dirichlet sectors, or simply using the bosonized expressions of these fields. The results are
\begin{equation}
\begin{split}
\langle\bar{\sigma}'_B(z_1)\sigma'&_B(z_2)\bar{\sigma}_B(z_3)\sigma_B(z_4)\rangle=\dfrac{1}{2z_{21}^{9/8}z_{43}^{1/8}}\left(\dfrac{z_{41}z_{32}}{z_{42}z_{31}}\right)^{3/8}\,,\\
&\langle\bar{\sigma}'_B(z_1)\sigma_B(z_2)\bar{\sigma}'_B(z_3)\sigma_B(z_4)\rangle=0\,.
\end{split}
\end{equation}

\cleardoublepage
\addcontentsline{toc}{section}{References}
\bibliography{bibliography}
\bibliographystyle{JHEP}

\end{document}